\documentclass[report]{JHEP3}
\usepackage{epsfig}
\title{DLCQ String Spectrum from ${\cal N}=2$ SYM Theory}
\usepackage{axodraw}

\author{Giuseppe De Risi\\Dipartimento di Fisica and Sezione
I.N.F.N., Universit\`a di Bari,Via G. Amendola 173, 70126
Bari, Italia. \email{E-mail:derisi$@$ba.infn.it}}

\author{Gianluca Grignani\\Dipartimento di Fisica and Sezione
I.N.F.N., Universit\`a di Perugia, Via A. Pascoli I-06123,
Perugia, Italia. \email{E-mail:grignani$@$pg.infn.it}
\thanks{Work supported in part by INFN and MURST of Italy.}}

\author{Marta Orselli\\Dipartimento di Fisica and Sezione
I.N.F.N., Universit\`a di Perugia, Via A. Pascoli I-06123,
Perugia, Italia. \email{E-mail:orselli$@$pg.infn.it}
\thanks{Work supported in part by INFN and MURST of Italy.}}

\author{Gordon W. Semenoff\\Pacific Institute for Theoretical Physics and\\
Department of Physics and Astronomy,
University of British Columbia,
6224 Agricultural Road, Vancouver, British Columbia V6T 1Z1 Canada.
\email{E-mail: gordonws$@$physics.ubc.ca}
\thanks{Work supported in part by NSERC of Canada and the PIMS String
Theory CRG.}}

\abstract{We study non planar corrections to the spectrum of operators in the
${\mathcal N}=2$ supersymmetric Yang Mills theory which are dual to
string states in the maximally supersymmetric pp-wave background with a {\em compact} 
light-cone direction. The existence of a positive definite discrete light-cone momentum
greatly simplifies the operator mixing problem. We give some examples where the 
contribution of all orders in non-planar diagrams can be found analytically.
On the string theory side this corresponds to finding the spectrum of a string
state to all orders in string loop corrections.}


\preprint{}

\begin{document}

\newcommand{\indup}[1]{_{\mathrm{#1}}}
\newcommand{\hypref}[2]{\ifx\href\asklfhas #2\else\href{#1}{#2}\fi}
\newcommand{\sfrac}[2]{{\textstyle\frac{#1}{#2}}}
\newcommand{\half}{\sfrac{1}{2}}
\newcommand{\quarter}{\sfrac{1}{4}}

%symbols
\newcommand{\Op}{\mathcal{O}}
\newcommand{\order}[1]{\mathcal{O}(#1)}
\newcommand{\eps}{\varepsilon}
\newcommand{\Lagr}{\mathcal{L}}
\newcommand{\superN}{\mathcal{N}}
\newcommand{\gym}{g_{\scriptscriptstyle\mathrm{YM}}}
\newcommand{\gtwo}{g_2}
\newcommand{\Tr}{\mathop{\mathrm{Tr}}}
\renewcommand{\Re}{\mathop{\mathrm{Re}}}
\renewcommand{\Im}{\mathop{\mathrm{Im}}}
\newcommand{\Li}{\mathop{\mathrm{Li}}\nolimits}
\newcommand{\cdott}{\mathord{\cdot}}
\newcommand{\singlet}{{\mathbf{1}}}

%brackets
\newcommand{\lrbrk}[1]{\left(#1\right)}
\newcommand{\bigbrk}[1]{\bigl(#1\bigr)}
\newcommand{\vev}[1]{\langle#1\rangle}
\newcommand{\normord}[1]{\mathopen{:}#1\mathclose{:}}
\newcommand{\lrvev}[1]{\left\langle#1\right\rangle}
\newcommand{\bigvev}[1]{\bigl\langle#1\bigr\rangle}
\newcommand{\bigcomm}[2]{\big[#1,#2\big]}
\newcommand{\lrabs}[1]{\left|#1\right|}
\newcommand{\abs}[1]{|#1|}

%eqnarray
\newcommand{\nn}{\nonumber}
\newcommand{\nln}{\nonumber\\}
\newcommand{\nl}{\nonumber\\&&\mathord{}}
\newcommand{\nle}{\nonumber\\&=&\mathrel{}}
\newcommand{\eq}{\mathrel{}&=&\mathrel{}}
\newenvironment{myeqnarray}{\arraycolsep0pt\begin{eqnarray}}{\end{eqnarray}\ignorespacesafterend}
\newenvironment{myeqnarray*}{\arraycolsep0pt\begin{eqnarray*}}{\end{eqnarray*}\ignorespacesafterend}

%%%%%%%%%%%%%%%%%%%%%%%%%%%%%%%%%%%%%%%%%%%%%%%%%%%%%%%%%%%%%%%%%%%%%%%%%%%
%\def\Tr{\mathop{\mbox{Tr}}}
%\def\nn{\nonumber}
%\newcommand\eqn[1]{eq.~(\ref{#1})}
%\newcommand{\binomial}[2]{\left ( \matrix{#1\cr #2\cr}\right )}
\newcommand{\ft}[2]{{\textstyle\frac{#1}{#2}}}
\newcommand{\cO}{{\cal O}}
\newcommand{\cT}{{\cal T}}
\def\ss{\scriptstyle}
\def\st{\scriptstyle}
\def\sst{\scriptscriptstyle}
\def\ra{\rightarrow}
\def\lra{\longrightarrow}
\newcommand\Zb{\bar Z}
\newcommand\bZ{\bar Z}
\newcommand\bF{\bar \Phi}
\newcommand\bP{\bar \Psi}

\newcommand{\be}{\begin{equation}}
\newcommand{\ee}{\end{equation}}
\newcommand{\bea}{\begin{eqnarray}}
\newcommand{\eea}{\end{eqnarray}}
\newcommand{\bino}[2]{\left( \begin{array}{c} #1 \\ #2 \end{array}
\right)}
\newcommand{\CW}{\mathcal{W}}
\newcommand{\CG}{\mathcal{G}}
\newcommand{\CO}{\mathcal{O}}
\newcommand{\CN}{\mathcal{N}}
\newcommand{\CI}{\mathcal{I}}
\newcommand{\D}{{\rm d}}
\newcommand{\ip}{{\rm i}}
\newcommand{\id}{\hbox{1\kern-.27em l}}
\newcommand{\sid}{\hbox{\scriptsize1\kern-.27em l}}
\newcommand{\pa}{\partial}
\newcommand{\rar}{\rightarrow}
\newcommand{\non}{\nonumber}
\newcommand{\we}{\kern-.1em\wedge\kern-.1em}
\newcommand{\scal}{\kern-.13em\cdot\kern-.13em}
\newcommand{\tF}{\tilde{F}}
\newcommand{\tA}{\tilde{A}}
\newcommand{\tE}{\tilde{E}}
\newcommand{\tB}{\tilde{B}}
\newcommand{\bLa}{{\bar{\La}}}
\newcommand{\bE}{{\bar{E}}}
\newcommand{\third}{\mbox{$\frac{1}{3}$}}
\newcommand{\fourth}{\mbox{$\frac{1}{4}$}}
\newcommand{\sixth}{\mbox{$\frac{1}{6}$}}
\newcommand{\II}{I\kern-.09em I}
\newcommand{\al}{\alpha}
\newcommand{\ga}{\gamma}
\newcommand{\Ga}{\Gamma}
\newcommand{\bet}{\beta}
\newcommand{\ka}{\kappa}
\newcommand{\de}{\delta}
\newcommand{\ep}{\epsilon}
\newcommand{\si}{\sigma}
\newcommand{\la}{\lambda}
\newcommand{\ta}{\tau}
\newcommand{\om}{\omega}
\newcommand{\Om}{\Omega}
\newcommand{\ze}{\zeta}
\newcommand{\De}{\Delta}
\newcommand{\La}{\Lambda}
\newcommand{\tha}{\theta}
\newcommand{\Ups}{\Upsilon}
\newcommand{\bi}{\bibitem}
\newcommand{\gt}{\tilde{g}}
\newcommand{\gwedge}{g^1 \wedge g^2 \wedge g^3 \wedge g^4}
\newcommand{\wg}{\wedge}
\newcommand{\del}{\partial}
\newcommand{\delbar}{{\bar\partial}}
\newcommand{\bz}{{\bar z}}
\newcommand{\tphi}{{\tilde\phi}}
\newcommand{\mC}{{\mathbf C}}
\newcommand{\mR}{{\mathbf R}}
\newcommand{\mZ}{{\mathbf Z}}
\newcommand{\mA}{{\mathbf A}}
\newcommand{\mB}{{\mathbf B}}
\newcommand{\mV}{{\mathbf V}}
\newcommand{\mU}{{\mathbf U}}
\newcommand{\Z}{\mathbb{Z}}
\newcommand{\R}{\mathbb{R}}
\newcommand{\nabsq}{\vec{\nabla}^2}
\newcommand{\ord}{\mathcal{O}}
\newcommand{\ads}{{\rm AdS}}
\newcommand{\spa}{\ \ ,\ \ \ \ }
\newcommand{\str}{\mathop{{\rm Str}}}
\newcommand{\tr}{\mathop{{\rm Tr}}}
\newcommand{\sn}{\mathop{{\rm sn}}}
\newcommand{\Ord}{{\cal{O}}}
\newcommand{\vecto}[2]{\left( \begin{array}{c} #1 \\ #2 \end{array}\right) }
\newcommand{\matrto}[4]{\left( \begin{array}{cc} #1 & #2 \\ #3 & #4
\end{array} \right) }
\newcommand{\calF}{\mathcal{F}}
\newcommand{\calL}{\mathcal{L}}
\newcommand{\calH}{\mathcal{H}}
\newcommand{\el}{\parallel}
\newcommand{\Hlc}{H_{\rm lc}}
\newcommand{\Hlct}{\tilde{H}_{\rm lc}}
\newcommand{\gqgt}{g_{\mathrm{QGT}}}
\newcommand{\map}{\cong}
\newcommand{\mapsim}{\sim}
\def\be{\begin{equation}}
\def\ee{\end{equation}}
\def\bea{\begin{eqnarray}}
\def\eea{\end{eqnarray}}
\def\nn{\nonumber}
\def\const{{\rm const}}
\def\v{\varphi}
\def\s {\sigma}
\def\a {\alpha}
\def\t {\tau}
\def\o {\omega}
\def\vcl{\varphi_{\rm cl}}
\newcommand{\no}[1]{:\!#1\!:}
\def\la{\left\langle}
\def\ra{\right\rangle}
\def\d{\partial}
\def\se{S_{\rm eff}}
\def\Oc{\mathcal{O}}
\def\Tr{\textrm{Tr}}
\def\Min{\textrm{Min}}
\def\Ab{\bar{A}}
\newcommand{\beq}{\begin{eqnarray}}% can be used as {equation} or {eqnarray}
\newcommand{\eeq}{\end{eqnarray}}
\newcommand{\bphi}{{\bf \Phi}}
\def\be{\begin{equation}}
\def\ee{\end{equation}}
\def\bea{\begin{eqnarray}}
\def\eea{\end{eqnarray}}
\def\trsm{{\rm\ Tr\,}}
\def\tr{{\bf\ Tr \,}}
\def\S{{\bf S}}
\newcommand{\newsection}[1]{
\vspace{10mm}
\pagebreak[3]
\addtocounter{section}{1}
\setcounter{subsection}{0}
\noindent
{\large\bf \thesection. #1}
\nopagebreak
\medskip
\nopagebreak}
\newcommand{\newsubsection}[1]{
\vspace{5mm}
\pagebreak[3]
\addtocounter{subsection}{1}
\addcontentsline{toc}{subsection}{\protect
\numberline{\arabic{section}.\arabic{subsection}}{#1}}
\noindent{\em %\thesection.
\thesubsection. #1}
\nopagebreak
\vspace{2mm}
\nopagebreak}
\setcounter{footnote}{1}
\newcommand{\figuren}[3]{\addtocounter{figure}{1}
\begin{figure}[htb]\begin{center}
\leavevmode\hbox{\epsfxsize=#2 \epsffile{#1.eps}}\\[3mm] \bigskip
\parbox{14.5cm}{\small \bf Fig.\thefigure.\ \it  #3}\vspace{-5mm}
\end{center}\end{figure}}

\section{Introduction}

The AdS/CFT correspondence asserts an exact duality between a
ten-dimensional type IIB superstring theory on $AdS_5\times S^5$
background and ${\cal N}=4$ supersymmetric Yang-Mills theory in
flat four dimensional Minkowski spacetime \cite{Maldacena:1997re,Gubser:1998bc,
Witten:1998qj}.
Though it has many spectacular successes, it is still a conjecture
and it is not yet clear whether it is an exact correspondence, or
is only valid in some limits of the two theories.  Given its
potential importance as a quantitative tool for strongly coupled
gauge and string theory, it is important to check it wherever
possible.

AdS/CFT is a strong coupling -- weak coupling duality.  This makes
it powerful, as it can be used to compute the strong coupling
limit of either theory using the weak coupling limit of the other.
On the other hand, it makes it difficult to check since it is not
easy to find situations where approximate computations in both
theories have an overlapping domain of validity. Early exceptions
to this were some quantities which were protected by supersymmetry
and didn't depend on the coupling constant at all
\cite{Lee:1998bx}, or quantities determined by anomalies which had
a trivial dependence \cite{Freedman:1998tz,Chalmers:1998xr} and a
few others related to circular Wilson loops
\cite{Erickson:2000af,Drukker:2000rr,Semenoff:2001xp,Semenoff:2002kk}
which had a nontrivial dependence on the coupling constant and
which could be computed for all values of the coupling.

More recently it has been realized that some large quantum number
limits  yield domains where accurate computations in both gauge
theory and string theory could be done and compared directly with
each other. The first and most powerful of these is the BMN limit.
It began with the observation \cite{Blau:2002dy,Gueven:2000ru} that the Penrose
limit of the $AdS_5\times S^5$ metric and 5-form field strength of
the string background are the maximally supersymmetric pp-wave
metric and a constant self-dual 5-form
\begin{eqnarray}
ds^2=-4dx^+dx^-+\sum_{i=1}^8 dx^idx^i -
\sum_{i=1}^8(x^i)^2dx^+dx^+\label{metric}
\\F_{+1234}=F_{+5678}={\rm const.}
\label{5form}
\end{eqnarray}
respectively.  Then Metsaev \cite{Metsaev:2001bj} found an exact
solution of the non-interacting type IIB Green-Schwarz string in
the background (\ref{metric}) and (\ref{5form}). Shortly
afterward, Berenstein, Maldacena and Nastase (BMN)
\cite{Berenstein:2002jq} noted that one could take a similar limit
of ${\cal N}=4$ supersymmetric Yang-Mills theory by considering
states with large R-charge. They identified the Yang-Mills
operators (called BMN operators) corresponding to the free string
states on the pp-wave background.

The AdS/CFT correspondence predicts that the spectrum of scaling
dimensions and charges under R-symmetry of these operators in the
't Hooft planar limit \cite{'tHooft:1973jz} of Yang-Mills theory should
match the free string spectrum. The leading order Yang-Mills
theory computation in ref.\cite{Berenstein:2002jq} and a two loop
calculation in ref.\cite{Gross:2002su} showed beautiful agreement.

Non-planar corrections to the spectrum of operators in Yang-Mills
theory should correspond to string loop corrections in string
theory.  The question of non-planar corrections to the spectrum of
BMN operators was considered in refs.\cite{Kristjansen:2002bb} and
\cite{Constable:2002hw}.  It was found that, once operator mixing
by non-planar graphs was resolved
\cite{Bianchi:2002rw,Beisert:2002bb,Constable:2002vq}, Yang-Mills
theory could be used to make predictions for the spectrum of
string states on the pp-wave background.  There are still ongoing
attempts to check these predictions on the string side of the
correspondence using light-cone string field theory
\cite{Spradlin:2002ar,Chu:2002pd,Spradlin:2002rv,
Chu:2002qj,Chu:2002eu,Pankiewicz:2002gs,Pankiewicz:2002tg,Chu:2002wj,He:2002zu,Gutjahr:2004dv}.
Success or failure of this matching would be a highly nontrivial
test of the AdS/CFT correspondence at the level of interacting
strings. There have also been interesting non trivial tests of the AdS/CFT 
correspondence in the BMN limit at finite temperature~\cite{PandoZayas:2002hh}.

In this paper, we shall discuss a  generalization of the BMN limit
that was found by Mukhi, Rangamani and Verlinde
\cite{Mukhi:2002ck}.  They showed that a certain limit of an
orbifold of $AdS_5\times S^5$ gives the plane-wave geometry
(\ref{metric}) and (\ref{5form}) with the additional feature that
the null coordinate $x^-$ is identified periodically $x^-\sim
x^-+2\pi R^-$. The result is a discrete light-cone quantized
string theory on the plane wave background. This is a
generalization of the BMN limit where the light-cone momentum is
discrete and there are wrapped states. The gauge theory which is
dual to the orbifolded string theory is an ${\cal N}=2$
superconformal Yang-Mills theory. The operators of Yang-Mills
theory which are dual to the string states were identified in
ref.\cite{Mukhi:2002ck}. Some checks that loop diagrams in planar
Yang-Mills theory reproduce the correct spectrum and a number of
generalizations to other types of orbifolds and limits to obtain
other compactifications of the pp-wave were considered in
ref.\cite{Bertolini:2002nr}.

We shall study non-planar corrections to the spectrum of the
operators in the ${\cal N}=2$ supersymmetric Yang-Mills theory
which are dual to single and multi-string states. Our main
observation is that, the existence of a positive definite,
discrete light-cone momentum  greatly simplifies the operator
mixing problem. In Yang-Mills theory, string interactions show up
in the mixing by non-planar diagrams of single trace and
multi-trace operators, the counterparts of single and multi-string
states. Since, in the case that we shall consider, the light-cone
momentum must be conserved and is discrete and positive, the
number of operators with different traces which can mix at any
level turns out to be finite and diagonalizing their mixing
exactly is a finite problem. We use this observation to give some
examples where the contribution of all orders in non-planar
diagrams can be found analytically. On the string side, this
corresponds to finding the spectrum of a string state to all
orders in string loop corrections (and, since we are doing
Yang-Mills perturbation theory, to leading orders in world-sheet
momenta).

In the next Section, we fix the notation and give a brief review
of ref.\cite{Mukhi:2002ck}.

\section{Preliminaries}

In this Section, we will first describe the gauge and string
theories which are dual to each other.  Then we will discuss the
Penrose limit of the string theory and the equivalent double
scaling limit of the gauge theory.  We will discuss the
holographic dictionary of non-interacting string theory states and
single trace operators in the gauge theory.

\subsection{${\mathcal N}=2$  from ${\mathcal N}=4$ }

The four-dimensional ${\cal N}=2$ supersymmetric Yang-Mills theory
that we are interested in can be obtained from its parent ${\cal
N}=4$ theory by an orbifold projection \cite{Douglas:1996sw, Bershadsky:1998cb}.
We begin with ${\cal N}=4$ with a $U(MN)$ gauge group. The
orbifold group will be the cyclic group $Z_M$ whose generator
$\gamma$ acts on the six scalar fields of ${\cal N}=4$ theory as
\be
\gamma:\left(\frac{\phi^1+i\phi^2}{\sqrt{2}},\frac{\phi^3+i\phi^4}{\sqrt{2}},
\frac{\phi^5+i\phi^6}{\sqrt{2}}\right)= \left(\omega
\frac{\phi^1+i\phi^2}{\sqrt{2}},\omega^{-1}
\frac{\phi^3+i\phi^4}{\sqrt{2}},
\frac{\phi^5+i\phi^6}{\sqrt{2}}\right),~~~
\omega= e^{{2\pi i\over M}} 
\label{aledef} 
\ee At the same time, this
transformation is  embedded into the gauge group as the group
element
\begin{equation}
g =\left( \matrix{ 1 & 0 & 0 & ... & 0 \cr 0 & \omega & 0 & ... &
0 \cr 0 & 0 & \omega^2& ... & 0 \cr 0  & 0 & 0 & ... & ... \cr 0 &
0 & 0 & ...& \omega^{N-1}\cr }\right)\label{projection}
\end{equation}
where each block of this $N\times N$ matrix is tensored with the
$M\times M$ unit matrix. Then, some components of the $MN\times
MN$ matrix fields are set to zero so that the equation
$$
\Psi = g \left(\gamma:\Psi\right) g^{\dagger}
$$
is satisfied, for all fields $\Psi$ of the ${\cal N}=4$ theory.
The resulting ${\cal N}=2$ theory has residual R-symmetry
$U(1)\times SU(2)$ and gauge group
 \be U(N)^{(1)}\times U(N)^{(2)} \times\cdots
U(N)^{(M)}. \ee $U(N)^{(M+1)}$ is identified with $U(N)^{(1)}$.

The resulting field content is as follows:
\begin{itemize}

\item{}$M$ vector multiplets \begin{equation}(A_{\mu
I},\Phi_I,\psi_{I},\psi_{\Phi I})~~,~~ I=1,...,M.\end{equation}
$\Phi_I$ is a complex scalar field and the Weyl fermion
$\psi_{\Phi I}$ is its superpartner.  $A^\mu_I$ is the gauge field
and $\psi_I$ is the gaugino.  All of these fields transform in the
adjoint representation of $U(N)^{(I)}$.

\item{}$M$ bi-fundamental hypermultiplets which, in ${\mathcal
N}=1$ notation, are
\begin{equation}(A_I,B_I,\chi_{AI},\chi_{BI})\end{equation}
The complex scalar field $A_I$ and its super-partner $\psi_{AI}$
transform in the $({ N}_I,{ \bar N}_{I+1})$ representation of
$U(N)^{(I)}\times U(N)^{(I+1)}$. The pair $B_I$ and $\chi_{BI}$
transform in the complex conjugate representation $({ \bar
N}_I,N_{I+1})$.
\end{itemize}

The ${\cal N}=2$ action can be found from the ${\cal N}=4$ theory.
The Euclidean Lagrangian density of ${\cal N}=4$ is
\begin{equation}
\label{actn4} {\cal L}= \frac{1}{\gym^2}   \mbox{TR} \left(
\frac{1}{2} F_{\mu\nu} F_{\mu \nu}  +   D_\mu \phi^i D_\mu \phi^i
- \sum_{i<j} [\phi^i , \phi^j] [\phi^i , \phi^j] +  \bar{\Psi}
\Gamma^\mu D_\mu \Psi +   \bar{\Psi} \Gamma^i [ \phi^i , \Psi]
\right)
\end{equation}
All fields are $MN\times MN$ matrices. With the notation
\begin{equation}
\label{ws} \mathbf A = \frac{1}{\sqrt{2}} ( \phi^1 + i \phi^2 )
\spa\mathbf B = \frac{1}{\sqrt{2}} ( \phi^3 + i \phi^4 ) \spa
\bphi = \frac{1}{\sqrt{2}} ( \phi^5 + i \phi^6 ) ~.
\label{scalarnotation}\end{equation} the  elements of the bosonic
fields which survive the projection (\ref{projection}) are
 \be {\bphi} \equiv \pmatrix{\Phi_1 &0 &\cdots &0\cr 0 &\Phi_2
&\cdots &0\cr \vdots & &\ddots & \vdots\cr 0 &0 &\cdots
&\Phi_{M}\cr} \qquad {A_{\mu}} \equiv \pmatrix{A_{\mu 1} &0
&\cdots &0\cr 0 &A_{\mu 2}  &\cdots &0\cr \vdots & &\ddots &
\vdots\cr 0 &0 &\cdots &A_{\mu M}\cr} \ee and \be \mathbf A \equiv
\pmatrix{0 &A_1& 0&\cdots &0\cr 0& 0 &A_2 &\cdots &0\cr \vdots& &
& \ddots&\vdots\cr 0 & 0 & 0 &\cdots &A_{M-1}\cr A_{M}& 0 & 0
&\cdots &0\cr} \qquad \mathbf B \equiv \pmatrix{0 &0 & \cdots& 0
&B_{M}\cr B_1& 0 &\cdots &0 &0\cr 0 & B_2 & \cdots &0 &0\cr
\vdots& & \ddots &\vdots &\vdots\cr 0& 0 & \cdots &B_{M-1} &0\cr}
\label{ampab} \ee Each non-vanishing entry of the above matrices
is an $N \times N$ matrix and corresponds to a bosonic field of
the ${\mathcal N}=2$ theory.

The $\mathcal N=4$ spinor $\Psi$ contains four different complex
Weyl spinors ${\bf\chi_A}$,${\bf \chi_B}$,${\bf \psi_\Phi}$ and
$\psi$ so that, with the definition in (\ref{scalarnotation})
${\bf \chi_A}$, ${\bf \chi_B}$ and ${\bf \psi_{\Phi}}$ are the
superpartners of ${\bf A}$, ${\bf B}$ and ${\bf \Phi}$,
respectively, and $\psi$ is the gaugino. Then
\be {\mathbf{\psi_{\Phi}}} \equiv \pmatrix{\psi_{\Phi 1} &0
&\cdots &0\cr 0 &\psi_{\Phi 2}  &\cdots &0\cr \vdots & &\ddots &
\vdots\cr 0 &0 &\cdots &\psi_{\Phi {M}}\cr} \qquad \mathbf{{\psi}}
\equiv \pmatrix{\psi_1 &0 &\cdots &0\cr 0 &\psi_2  &\cdots &0\cr
\vdots & &\ddots & \vdots\cr 0 &0 &\cdots &\psi_{M}\cr} \ee and
\be \mathbf{\chi_A} \equiv \pmatrix{0 &\chi_{A1}& 0&\cdots &0\cr
0& 0 &\chi_{A2} &\cdots &0\cr \vdots& &  & \ddots&\vdots\cr 0 & 0
& 0 &\cdots &\chi_{A{M-1}}\cr \chi_{A{M}}& 0 & 0 &\cdots &0\cr}
\mathbf{\chi_B} \equiv \pmatrix{0 &0 & \cdots& 0 &\chi_{B{M}}\cr
\chi_{B1}& 0 &\cdots &0 &0\cr 0 & \chi_{B2} & \cdots &0 &0\cr
\vdots& & \ddots &\vdots &\vdots\cr 0& 0 & \cdots &\chi_{B{M-1}}
&0\cr} \label{psichis} \ee An element of the residual gauge group
is \be U \equiv \pmatrix{U^{(1)} &0 &\cdots &0\cr 0 &U^{(2)}
&\cdots &0\cr \vdots & &\ddots & \vdots\cr 0 &0 &\cdots
&U^{(M)}\cr} \ee and acts on all of the above matrices by
conjugation, $M\to UMU^{\dagger}$ (with appropriate additional
terms for the gauge field).

The action for scalar fields is \bea {\cal L}_{\rm scalar}=
\frac{1}{\gym^2M}\sum_{I=1}^M \mbox{TR} \left( D_\mu \mathbf A_I
D_\mu \bar{\mathbf A}_I + D_\mu \mathbf B_I D_\mu \bar{\mathbf
B}_I + \frac{1}{2} D_\mu \bphi_I D_\mu \bar{\bphi}_I \right) +
\mathcal{L}_D + \mathcal{L}_F\spa \eea
where the interaction $F$- and $D$-terms can be gotten from
\bea \mathcal{L}_D &= - \frac{1}{\gym^2M}  \mbox{TR} \left(
[\mathbf A,\bar{\mathbf A}] + [\mathbf B,\bar{\mathbf B}] +
[\bphi,\bar{\bphi}] \right)^2 \spa \label{Dterms}\\
\mathcal{L}_F &= \frac{2}{\gym^2M} \mbox{TR} \left( \left|
[\mathbf A,\mathbf B] \right|^2 + \left| [\mathbf A,\bphi]
\right|^2 + \left| [\mathbf B,\bphi] \right|^2
\right)~.\label{Fterms} \eea

The factor of $1/M$ is the order of the orbifold group and it
comes from the orbifold projection.

%The matter content of the gauge theory is well represented by a
%quiver/moose diagram~\cite{Mukhi:2002ck}, see Fig.1.
%
%\figuren{largemoose}{7cm}{Each line connecting nodes corresponds to the
%bifundamental
%fields $A_I$ and $B_I$ ($I=1,\dots, M$), the two lines together
%correspond to the
%bifundamental hypermultiplet. The line going back to the same node
%represents the adjoint scalars $\Phi_I$. The arrows go from fundamental to
%anti-fundamental representation of the corresponding gauge groups.}

%In Ref.s~\cite{Bershadsky:1998mb,Bershadsky:1998cb} it was shown
%that the correlation functions of the ${\mathcal N}=2$ theory are
%the same as those of the corresponding ${\mathcal N}=4$ SYM theory
%in the planar limit, provided $\gqgt^2 = \gym^2 M = 4 \pi g_s M$
%are identified. This relation also follows from taking
%(\ref{actn4}) as the action for ${\mathcal N}=2$ QGT.

\subsection{IIB String on $AdS_5\times \S^5/\mZ_{M}$}

The ${\cal N}=2$ theory is the holographic dual of IIB string
theory with background the orbifold $AdS_5\times \S^5/\mZ_{M}$ and
with $MN$ units of Ramond-Ramond 5-form flux through the 5-sphere.
Since the 5-sphere contains $M$ copies of a fundamental domain
that are identified by the orbifold group, there are $N$ units of
flux per fundamental domain. The action of the orbifold group is
obtained by embedding the 5-sphere in $\mR^6\sim \mC^3$ so that
$$ \sum_{i=1}^3\left| z_i\right|^2=R^2
$$ where $(z_1,z_2,z_3)\in \mC^3$ and then identifying points as
prescribed in (\ref{aledef}): \begin{equation} \left(
z_1,z_2,z_3)\sim (\omega
z_1,\omega^{-1}z_2,z_3\right)~~.~~\omega=e^{2\pi i/M}
\label{ident}
\end{equation}

 The radii of $AdS_5$ and $\S^5$ are equal and are
given by \be R^2  = \sqrt{4\pi g_s \alpha'^2 N M} \ ,
\label{adsradius} \ee where $g_s$ is the type IIB string coupling.
Furthermore, the Yang-Mills theory coupling constant of the parent
${\cal N}=4$ theory is identified with the coupling constant of
the parent superstring theory on $AdS_5\times \S^5$,
\begin{equation}
4\pi g_s=g_{YM}^2
\end{equation}

\subsection{Double Scaling limit}

We shall consider the double scaling limit of both the gauge
theory and its string theory dual.   The double scaling limit of
the string theory is the Penrose limit which obtains the pp-wave
background. The radii of $AdS_5$ and $ \S^5 $, given by $R$ in
(\ref{adsradius}), are put to infinity by scaling both $N$ and $M$
to infinity while keeping $g_s$ small but finite. The parameter
which will become the null compactification radius, $R^-={R^2\over 2 M}$,
is also held fixed in the limit by keeping the ratio $N\over M$
fixed.

The metric of $AdS_5\times \S^5/\mZ_{M}$ can be written as:
\bea
ds^2 &=& R^2\Bigg[-\cosh^2\rho \, dt^2 + d\rho^2 + \sinh^2\rho\,
d\Omega_3^2 + \nonumber\\
&& d\alpha^2 + \sin^2\alpha\, d\theta^2 + {\cos^2\alpha}\,
\left(d\gamma^2 +
\cos^2\gamma\, d\chi^2+\sin^2\gamma\, d\phi^2\right)\Bigg]
\ .
\label{adsmet}
\eea
  The
angles of $\S^5$ are related to the complex coordinates of
$\mC^3/\mZ_{M}$ by \be   z_1=R\cos\alpha \cos\gamma\, e^{i\chi},
\quad z_2=R\cos\alpha \sin\gamma\, e^{i\phi}\,,~~z_3=R \sin\alpha
e^{i\theta} \ee In terms of the angles of $\S^5$ the orbifold
described by the action (\ref{ident}) is obtained by  the
identifications \be \chi\sim \chi+{2\pi \over M},\qquad \phi\sim
\phi-{2\pi\over M} \ . \label{periodic} \ee

To take the Penrose limit it is useful to introduce the
coordinates \bea &r=\rho R,~~w= \alpha R, & y=\gamma R \ .
\label{pplimit} \eea and the light-cone coordinates \bea & x^+ =
\frac12\left(t+\chi\right), & x^- = {R^2\over
2}\left(t-\chi\right) \label{lcone} \ . \eea After taking the
$R\rightarrow \infty$ limit and renaming some coordinates, the
metric becomes
 \cite{Alishahiha:2002ev}

\be ds^2 = -4dx^+ dx^- - \sum_{i=1}^8 (x^i)^2 \,{dx^+}^2 +
\sum_{i=1}^8 {dx^i}^2 \ , \label{ppwaveg} \ee In the geometry
(\ref{ppwaveg}) there is also a Ramond-Ramond flux \be F_{+1234} =
F_{+5678} = const \ . \label{rrflx} \ee So far, with the rescaling
(\ref{pplimit}) and (\ref{lcone})  the only limit that we have
taken to obtain (\ref{ppwaveg}) is that of large $R$. The orbifold
identification (\ref{periodic}) implies that the light-cone
coordinates have the periodicity \bea
&& x^+\sim x^+ + {\pi\over M}\nonumber\\
&& x^- \sim x^- + {\pi R^2\over M} \ , \eea   In the double
scaling limit, as $R$ is taken large, $M$ is also taken large so
that $R^-={R^2\over 2M}$ is held fixed. In the limit
\begin{equation}
\left( x^+,x^-\right)\sim\left( x^+,x^-+2\pi R^-\right)
\end{equation}
The periodic direction becomes null. As a consequence the
corresponding light-cone momentum $2p^+$ is quantized in units of
$1\over R^-$.

The conclusion is that the Penrose limit of $AdS_5\times
\S^5/\mZ_{M}$ with $M\to \infty$  in this particular way leads to
a Discrete Light-Cone Quantization (DLCQ) of the string on a
pp-wave background, in which the null coordinate $x^-$ is
periodic. Note that the orbifold of the 5-sphere preserves half of
the supersymmetries of the original $AdS_5\times \S^5$ solution of
string theory. Nonetheless, in the Penrose limit, we recover the
maximally supersymmetric plane-wave background.

Discrete light-cone quantization of the string on the pp-wave
background is a slight generalization of
ref.\cite{Metsaev:2001bj}.  One component of the light-cone
momentum is quantized as
\begin{equation}
2p^+=\frac{k}{R^-}  ~~,~~ k=1,2,3,...
\end{equation}
The other component is the light-cone-gauge Hamiltonian,
\begin{eqnarray}  2p^- &=& \sum_{n = -\infty}^{\infty}
\,\left(\sum_{i=1}^8a_n^{i\dagger}a^i_n+\sum_{\alpha=1}^8
b_n^{\alpha\dagger}b_n^\alpha\right)\sqrt{1
+ {4n^2
(R^-)^{2}\over k^2 \alpha'^2 }}\nonumber \\
&=&\sum_{n = -\infty}^{\infty}
\,\left(\sum_{i=1}^8a^{i\dagger}_na^i_n+\sum_{\alpha=1}^8
b_n^{\alpha\dagger}b_n^\alpha\right)\sqrt{1 + {4\pi g_s N\over
M}\frac{n^2}{k^2}} \label{stringspectrum}\end{eqnarray} where
$a^i_n,a^{i\dagger}_n$ and $b_n^{\alpha},b_n^{\alpha\dagger}$ are
the annihilation and creation operators for the discrete bosonic
and fermionic transverse oscillations of the string, respectively.
They obey the (anti-) commutation relation
\begin{equation} \left[ a^i_{n_1},
a^{j\dagger}_{n_j}\right]=\delta^{ij}\delta_{n_in_j}~~,~~ \left\{
b^\alpha_{n_1},b^{\beta\dagger}_{n_j}\right\}=\delta^{\alpha\beta}\delta_{n_in_j}\end{equation}
In the last line of eqn.(\ref{stringspectrum}) we have written the
compactification radius in terms of string background parameters.

There are also wrapped states. If the total number of times that
the closed string wraps the compact null direction is $m$, the
level-matching condition is
\begin{equation} km = \sum_{n= -\infty}^{\infty}\, n \left(\sum_{i=1}^8
a^{i\dagger}_na^i_n +\sum_{\alpha=1}^8
b_n^{\alpha\dagger}b_n^\alpha\right)\ , \label{sspectrum}
\end{equation}

States of the string are characterized by their discrete
light-cone momentum $k$ and their wrapping number $m$.  The lowest
energy state in a given sector is the string sigma model vacuum,
$\left| k,m\right>$ which obeys $$ a_n^i\left| k,m\right>
=0=b_n^\alpha\left| k,m\right> ~~,~~\forall n,i,\alpha$$ Other
string states are built from the vacuum by acting with transverse
oscillators,
  \be \prod_{j = 1}^{L} \, a^{i_j\dagger}_{n_j} \prod_{j' =
1}^{L'} \, b^{\alpha_{j'}\dagger}_{n_{j'}} \mid k,m\rangle
\label{stgst} \ee The level matching condition reads \be
 \sum_{j =1 }^{L} \, n_j +\sum_{j'=1}^{L'}\, n_{j'}= k\,m.
\label{momcons}
\ee

\subsection{Matching charges}

There are three important quantum numbers that can be matched
between the string theory and its gauge theory dual.   One is the
energy in string theory, which is the quantum operator generating
a flow along the Killing vector field $i\partial_t$ of the
background.  It corresponds to the conformal dimension, $\Delta$,
of operators in the gauge theory.

The others are U(1) charges.  Two are particularly important to
us. One is $J'$ which generates a U(1) which is in the SU(2)
subgroup of the R-symmetry
$$
A\to e^{i\xi}A~~,~~ B\to e^{i\xi}B~~,~~0\leq\xi<2\pi
$$
$J'$ which has integer eigenvalues. In the orbifold geometry, it
corresponds to the Killing vector
$J'=-\frac{i}{2}\left(\partial_\chi+\partial_\phi\right)$.  There
is an additional $U(1)$ which is not part of the R-symmetry
$$
A\to e^{i\zeta}A~~,~~B\to e^{-i\zeta}B~~,~~0\leq\zeta<2\pi/M
$$
The domain of the angle $\zeta$ is reduced from $2\pi$ to $2\pi/M$
by the orbifold identification.  This $U(1)$ is generated by $J$
whose eigenvalues are integer multiples of $M$. In order to
normalize it more conveniently, we rename it $MJ$ where $J$ has
integer eigenvalues. On the orbifold geometry, it corresponds to
the Killing vector
$J=-\frac{i}{2M}\left(\partial_\chi-\partial_\phi\right) $.

In summary, charges and Killing vectors are related by
$$
\Delta=i\partial_t
~~,~~J=-\frac{i}{2M}\left(\partial_\chi-\partial_\phi\right) ~~,~~
J'=-\frac{i}{2}\left(\partial_\chi+\partial_\phi\right)
$$
We can then recall the combinations of $\chi$, $\phi$ and $t$
which were used to form the light-cone coordinates $x^+$ and $x^-$
of the pp-wave geometry to deduce the light-cone momenta  \bea
2p^- =& {i} (\partial_t +\partial_\chi) ~=& \Delta
- M J -J'  \nonumber\\[2mm]
2 p^+ =& \displaystyle i {(\partial_t - \partial_\chi)\over R^2}
~=& {\Delta + M J +J' \over 2MR^-} \ .
\label{pplusminus}
\eea
These are the light-cone momenta of string states.  We will focus
on those states of the gauge theory where these quantum numbers
remain finite in the double scaling limit. It will be easy to see
that $2p^+$ will turn out to be quantized appropriately in units
of integers$/2R^-$ and the values of $2p^-$ which we find in the
gauge theory will be compared to the spectrum of the string
light-cone Hamiltonian.

The BPS condition $\Delta\geq \left|MJ+J'\right|$ implies that
keeping $2p^+$ and $2p^-$ finite as $R,M\to\infty$ will clearly
only be possible
  when both
$\Delta$ and $MJ+J'$ diverge with their difference,
$\Delta-(MJ+J')$, remaining finite.

The charges of gauge theory operators are obtained as follows. By
convention, the U(1) transformation is generated by $e^{4\pi iJ}$
so the $A_I$ and $B_I$ fields that make up the hypermultiplets
have fractional charge under $J$, ${1\over 2M}$ and $-{1\over 2M}$
respectively. The operator $J'$ generates a $U(1)$ symmetry
contained in the $SU(2)_R$ factor of the R-symmetry. Under this
$U(1) \subset SU(2)_R$, the fields $\Phi_I$ are neutral. On the
other hand, the scalars $A_I, B_I$ in the hypermultiplets both
have charge $\frac12$ under $J'$. Complex conjugation and
supersymmetry give the remaining charge assignments, for the
fermions and all the conjugate fields.

\begin{table}[t]
\begin{center}
\begin{tabular}[t]{c|c|c|c|c|p{1cm} c|c|c|c|c|}
~ & $\Delta$ & $M J$ & $J'$ & $2p^-$ & & & $\Delta$ & $M J$ & $J'$
& $2p^-$ \\[1mm]
\cline{1-5}\cline{7-11} &&&&&&&&&& \\[-4mm]
$A_I$ & 1 & ${1\over 2}$ &  ${1\over 2}$ & 0 && ${\overline A_I}$
& 1 & $-{1\over 2}$ & $-{1\over2}$ &
2 \\[1mm]
\cline{1-5}\cline{7-11} &&&&&&&&&& \\[-4mm]
$B_I$ & 1 & $-{1\over 2}$ &  ${1\over 2}$ &  1 && ${\overline
B_I}$ & 1 & ${1\over 2}$ & $-{1\over2}$ &
1 \\[1mm]
\cline{1-5}\cline{7-11} &&&&&&&&&& \\[-4mm]
$\Phi_I$ & 1 & 0 &  0  &  1 && ${\overline \Phi_I}$ & 1 & 0 &  0
&
1 \\[1mm]
\cline{1-5}\cline{7-11} &&&&&&&&&& \\[-4mm]
${\chi_{A}}_I$ & ${3\over 2}$ & ${1\over 2}$ &  0 &  1 &&
${\overline{\chi}_A}_I$ & ${3\over 2}$ & $-{1\over 2}$ &
0 & 2 \\[1mm]
\cline{1-5}\cline{7-11} &&&&&&&&&& \\[-4mm]
${\chi_{B}}_I$ & ${3\over 2}$ & $-{1\over 2}$ &  0 &  2 &&
${\overline{\chi}_B}_I$ & ${3\over 2}$ & ${1\over 2}$ &  0
& 1 \\[1mm]
\cline{1-5}\cline{7-11} &&&&&&&&&& \\[-4mm]
${\psi_\Phi}_I$ & ${3\over 2}$ & 0 &  $-{1\over 2}$ & 2 &&
${\overline {\psi}_\Phi}_I$ & ${3\over 2}$ & 0 &
${1\over 2}$ & 1 \\[1mm]
\cline{1-5}\cline{7-11} &&&&&&&&&& \\[-4mm]
$\psi_I$ & ${3\over 2}$ & 0 & $-{1\over2}$ & 2 &&
$\overline\psi_I$ & ${3\over 2}$ & 0 &
${1\over 2}$ & 1 \\[1mm]
\cline{1-5}\cline{7-11} \multicolumn{5}{c}{\vtop{\smallskip
\hbox{\small Table 1: Dimensions and charges for}
\smallskip
\hbox{~~~~~~~~~~~\small chiral fields and gauginos}}} &
\multicolumn{6}{c}{~~~~~~~~~\vtop{\smallskip\hbox{\small Table 2:
Dimensions and charges
for}\smallskip\hbox{~~~~~~~~~~~\small complex conjugate fields}}} \\
\end{tabular}\\
\end{center}
\end{table}

The dimension and charge assignments, along with the $2p^-$
values, are summarized in Tables 1 and 2. In Table 1, $A_I, B_I$
refer to the scalar components of the ${\cal N}=1$ chiral
superfields that form the ${\cal N}=2$ hypermultiplets.
$\chi_{A_I},\chi_{B_I}$ are their fermionic partners. $\Phi_I$ are
the complex scalars in the vector multiplet, while ${\psi_\Phi}_I$
are their fermionic partners. Finally, $\psi_I$ are the gauginos
in the theory. Table 2 lists the complex conjugate fields.

\subsection{The holographic dictionary}

In order to identify states in the ${\cal N}=2$ gauge theory with
finite values of light-cone momenta, as given in
(\ref{pplusminus}), we first find the appropriate quantum numbers
of the field operators.  These are tabulated in Table 1 and Table
2. We see that only the fields $A_I$ carry vanishing $2p^-$.  By
matching quantum numbers, we see that the string state $|k,0>$
corresponds to the gauge invariant composite operator
$$
\left|k,0\right>~~\leftrightarrow ~~{\rm TR}\left(
\left(A_1(x)A_2(x)\ldots A_M(x)\right)^k\right)
$$
We have indicated the $x$-dependence of the composite operator. In
the following, where from the context it is obvious, we will omit
it.  Because $A_I(x)$ transforms in the bi-fundamental
representation of the gauge group, we are required to form the
chains $A_1\ldots A_M$ to obtain a gauge invariant operator. This
chain can be repeated $k$ times. The conformal dimension of this
composite operator is protected by supersymmetry.   This
protection is inherited from the parent ${\cal N}=4$ theory. Thus,
its exact conformal dimension is $\Delta=km$ and its exact
spectrum is therefore $p^-=0$.

We have  chosen to use a one-trace operator to represent the
single string state.   Indeed, this choice has some arbitrariness.
A more general operator would be any linear combination of
multi-trace operators,
\begin{equation}
{\cal O}(\ell_1,\ell_2,...)=
 \left( {\rm
TR}\left( A_1\ldots A_M\right)\right)^{\ell_1} \left( {\rm
TR}\left( A_1\ldots A_M\right)^2\right)^{\ell_2} \left( {\rm
TR}\left( A_1\ldots A_M\right)^3\right)^{\ell_3}\ldots
\label{moregeneral}
\end{equation} 
where
$$
\sum_i \ell_i=k
$$
 Operators with different trace
structures are not mixed in the planar limit of Yang-Mills theory,
but non-planar corrections do mix them, though the mixing vanishes
in the double scaling limit. One natural way to decide which
amongst the degenerate states are relevant is to diagonalize the
inner product $\left<\ell_1,...|\ell_1',...\right>$ which one
would obtain from the correlation function
$$
\left<\bar{\cal O}(x;\ell_1,\ell_2,...){\cal
O}(y;\ell_1',\ell_2',...)
\right>=\frac{1}{(x-y)^{2k}}\left<\ell_1,...|\ell_1',...\right>
$$
However, there is no natural way to decide which of the resulting
states is a one-string state, two-string state, etc. This is
similar to the problem on the string side of trying to distinguish
the multi-string states $$
\left|\ell_1,0\right>\otimes\left|\ell_2,0\right>\otimes\left|\ell_3,0\right>
\otimes... $$ which, when $\sum \ell_i=k$, all have the same
quantum numbers.  At this point, this should be regarded as an
open problem. Fortunately, we shall find that we do not have to
solve this problem  here since we are interested in the
eigenvalues of the Hamiltonian which are independent of the basis.
We already know that these states are degenerate and have
eigenvalue $2p^-=0$.

There are eight states which are created by one bosonic oscillator
and eight which are created by a fermionic oscillator.  These all
add one unit to the Hamiltonian, $2p^-$.  In Yang-Mills theory,
they are gotten by inserting an impurity into the $A_1...A_M$
chains.  From Tables 1 and 2, we see that four of the bosonic
states are gotten by inserting $B_I$, $\Phi_I$, $\bar B_I$ or
$\bar \Phi_I$. The other four are gotten by replacing $A_I$ by a
derivative of $A_I$.  For example, a state with
$2p^-=1+$corrections is
\begin{equation}
\left(a^{5\dagger}_n+ia^{6\dagger}_
n\right)\left|k,m\right>~~\leftrightarrow~~\sum_{I=1}^{kM}e^{2\pi
i nI/kM}{\rm TR}\left( A_1\ldots A_{I-1}\Phi_IA_I\ldots
A_M\left(A_1\ldots A_M\right)^k\right) \label{1imp}\end{equation}
We have superposed over positions at which the impurity could be
inserted. The momentum in the insertion $n$ coincides with the
world-sheet momentum of the oscillator state.  The level matching
condition comes from realizing that the actual periodicity of the
operator is $I\to I+M$, rather than $I\to I+kM$, which the plane
waves anticipate. This requires that $n=km$, where $m$ is an
integer. This is the level matching condition.  The integer $m$ is
identified with the wrapping number of the world-sheet on the
compact coordinate.

The single oscillator state in (\ref{1imp}) is no longer a
protected operator.  Its dimension $\Delta$ should get radiative
corrections beyond the tree level in Yang-Mills theory, even for
planar diagrams.  In fact, it must get such corrections if it is
to match the string spectrum,
\begin{equation}
2p^-=\sqrt{ 1+\frac{g_{YM}^2N}{M}\frac{n^2}{k^2}} \label{1impspec}
\end{equation}
for planar diagrams.  Indeed, we shall see in the following that
it produces this spectrum to one order in $g_{YM }^2$.  We will
also learn that this operator is quasi-protected in that, in the
double scaling limit, all non-planar corrections to
(\ref{1impspec}) vanish. {\bf Our Yang-Mills computation predicts
that the spectrum of this state in string theory does not receive
string loop corrections.}

Here, one might wonder why, rather than (\ref{1imp}) we couldn't
insert one impurity into a multi-trace operator.  Indeed, when
$k>1$ there are multi-trace operators which have the same $k$ and
$m$ and which non-planar diagrams mix with (\ref{1imp}).
Moreover, since this mixing vanishes in the double scaling limit,
all such operators with the same $m$ and $k$ are exactly
degenerate. Again, one could choose a special basis by
diagonalizing their inner product, but we shall not have to do
this, as here we are interested only in questions about the
spectrum which are basis independent.

The winding state with two impurities which corresponds to two
oscillator states and have energies $2p^-=2+$corrections reads
\bea 
\left(a^{5\dagger}_{n_1}+ia^{6\dagger}_
{n_1}\right)\left(a^{5\dagger}_{n_2}+ia^{6\dagger}_
{n_2}\right)\left|k,m\right>~\leftrightarrow~
~~~~~~~~~~~~~~~~~~~~~~~~~~~~~~~~~~~~~~~~~~~~~~~~~~~~\nonumber
\\ 
\sum_{I,J=1}^{k M}{\rm TR}\left[A_1\ldots A_{I-1} \Phi_I A_I\ldots
A_{J-1} \Phi_J A_J\ldots A_M \left(A_1\ldots
A_M\right)^{k-1}\right] e^{2\pi i\frac{I n_1+J n_2}{Mk}}
\label{excstatek} \eea Here, the world-sheet momenta are $n_1$ and
$n_2$.  Also, note that the state is periodic under translating
both $I$ and $J$ by $M$.  This leads to the quantization condition
$n_1+n_2=k m$.  We interpret this as the level matching condition
and $m$ is the winding number.

In most of the above discussion, we have focused on the oscillators
constructed out of $\Phi$.  However, it is straightforward to see
that similar expressions hold for the remaining oscillators, with
$\Phi$ suitably replaced by the other type of impurities
or one of the fermionic fields.

\section{Spectrum of Strings from Yang-Mills Theory}

It is by now well known that composite operators made from gauge
invariant products of adjoint (and in our case bi-fundamental)
fields, have some special properties. For example, consider a
composite made from the scalar fields of ${\cal N}=4$
supersymmetric Yang-Mills theory,
$$ {\rm TR}\left( \phi^{i_1}\phi^{i_2}\ldots\phi^{i_k}\right) $$
In four space-time dimensions, the engineering dimension of a
scalar field is one and therefore, for all choices of the indices,
$i_1,\ldots,i_k$, the composite operators above have the same tree
level dimension.

Of course, at the quantum level, when radiative corrections are
taken into account, the degeneracy between such operators can be
lifted. Generally, loop corrections can be separated into two
different kinds.  One type are ``flavor blind'', they do not
distinguish between the different flavors, labelled by indices
$i_1,\ldots,i_k$, but only proceed through the fact that the
fields carry charges which couple to the gauge field. These
corrections provide the same overall constant shift in the
spectrum of all of the  operators and do not resolve the
degeneracy between them. In theories with enough supersymmetry,
these corrections can cancel identically. This is indeed the case
in the parent ${\cal N}=4$ theory and in the ${\cal N}=2$ theory
of interest.  An example of interactions which contribute to these
radiative corrections are the scalar four-point couplings in the
D-terms in (\ref{Dterms}). When combined with gauge field loops,
the interactions from D-terms cancel identically.

The other kind of radiative corrections do distinguish between
different flavors. They can split the tree level degeneracy of
conformal dimensions.   A well-known example occurs in the ${\cal
N}=4$ gauge theory where the F-terms couple the tree level
degenerate operators and act effectively like the integrable
Hamiltonian for an $SO(6)$ spin chain \cite{Minahan:2002ve}.
Again, in the example of interest to us, the ${\cal N}=2$ theory,
the F-terms (\ref{Fterms}) in the scalar four-point couplings also
split the spectrum of conformal dimensions. It can be shown that
they account for the entire radiative correction to one loop order
for the conformal dimensions of products of scalar fields that we
shall consider here.  The contribution is analyzed in Appendix A.

Just as in  ${\cal N}=4$ supersymmetric Yang-Mills theory
~\cite{Beisert:2002ff,Beisert:2003tq}, the computation of
anomalous dimensions is elegantly summarized by the action of an
effective Hamiltonian. In Appendix A we show that the one-loop
shift in dimension of all composite operators which have a certain
property can be summarized by the action on traces of matrices of
the dilatation operator
\begin{eqnarray}
D =\sum_{I=1}^M{\rm TR}\left(A_I\bar A_I+ B_I\bar
B_I+\Phi_I\bar\Phi_I\right)-\frac{g_{YM}^2M}{8\pi^2}
\sum_{I=1}^M{\rm TR}\left[A_I\Phi_{I+1} \bar A_I \bar \Phi_{I}
-\right.\nonumber \\ \left.- A_I\Phi_{I+1} \bar\Phi_{I+1} \bar A_I
 -\Phi_{I}A_I
\bar A_I\bar \Phi_{I} + \Phi_{I} A_I \bar\Phi_{I+1}\bar
A_I\right]+\ldots \label{dilatationoperator}
\end{eqnarray} 
The
terms with dots contain other matrices such as $B_I$ and $\bar
B_I$ and fermions which we will not use here. The operatorial
property is defined by the Wick contractions
$$
\left<\left[\bar A_I\right]_{ab}
\left[A_{J}\right]_{cd}\right>_0=\delta_{IJ}\delta_{ad}\delta_{bc}
~~,~~ \left<\left[\bar
B_I\right]_{ab}\left[B_{J}\right]_{cd}\right>_0=\delta_{IJ}\delta_{ad}\delta_{bc}
~~,~~ \left<\left[\bar \Phi_I\right]_{ab}
\left[\Phi_{J}\right]_{cd}\right>_0=\delta_{IJ}\delta_{ad}\delta_{bc}
$$

An example of a basis of operators (with one impurity and $k=2$)
is the set of $2M$ elements
$$
{\rm TR}\left( A_1\ldots A_{I-1}\Phi_IA_I...A_MA_1...A_M\right)
~~,~~ {\rm TR}\left( A_1\ldots A_{I-1}\Phi_IA_I...A_M\right){\rm
TR}\left(A_1...A_M\right)
$$
for each value of $I=1,...,M$. To operate on a such basis with $D$
we Wick-contract $\bar A_I$ and $\bar\Phi_I$ which occur in $D$
with each of the $A_I$ and $\Phi_I$ which occur in the operators,
respectively. In each case this produces a linear combination of
operators in the basis. The first term in
(\ref{dilatationoperator}) gives the tree level contribution to
the conformal dimension and the second term gives the one loop
correction.

Generally, there is a basis of operators $\Op_\alpha$ determined
by the quantum number $k$ and the number and types of impurities
and
$$
D \Op_\alpha = D_\alpha^{~\beta}\Op_\beta
$$
The eigenvectors of the matrix $D_\alpha^{~\beta}$ are the
scaling operators and the eigenvalues are the conformal
dimensions.

One immediate result is that the set of operators of the type
(\ref{moregeneral}), since they contain only $A_I$'s, do not get
corrections at all.  They are related to chiral primary operators
of ${\cal N}=4$ supersymmetric Yang-Mills theory and share that
property.  For them, $D$ is diagonal and its eigenvalue is $kM$.

In the next sections we will diagonalize the matrix
$D_{\alpha}{}^{\beta}$ for operators with one impurity and any
value of the discrete light cone momentum (which, as we shall
show, do not have an anomalous dimension beyond the planar level)
and  for operators with two impurities and the first few values of
the discrete light cone momentum $k=1,2,3$. Except for $k=1$ the
eigenstates of the dilatation matrix will not be eigenstates of
the winding number $m$ but will be linear combinations of these
eigenstates with different values of $m$.

The string spectrum can be obtained exactly in the large $M$ limit
where the variable $x=I/M$ becomes continuous and the action of
the Hamiltonian on states   is described by a simple differential
operator whose eigenstates and eigenvalues can be easily found. We
shall also make some comments on possible extensions of our
results to the case of higher values of $k$ and a larger number of
impurities.

\section{One impurity = one oscillator states}

Let us begin by considering the simplest states, those
which have $k=1$.  With no impurities, we already know that this
state has $2p^-=0$.   With one impurity it is convenient to take
the linear combination   \be {\cal {O}}_{n,1}=\sum _{I=1}
^{M}O^1_I \omega^{nI} \label{1impu} \ee where \be O^1_I= {\rm TR}
(A_1 A_2\cdots A_{I-1} \Phi_I A_I \cdots A_{M})\\
\label{1impk=1} \ee Here $\omega = e^{2\pi i\over M}$ and since
$k=1$, the level matching condition, $n=mk=m$, is trivially obeyed
when we identify $m$ as both the world-sheet momentum and the
wrapping number.  (\ref{1impk=1}) is to be treated as periodic in
$I$ with period $M$.  This has been anticipated in the Fourier
transform (\ref{1impu}).

The action of the dilatation operator (\ref{dilatationoperator})
on  (\ref{1impu}) is found by performing Wick contractions.
  \bea D\circ {\cal{O}}_{n,1}&=&(M+1){\cal
O}_{n,1}-\frac{ g_{YM}^2M N}{8\pi^2}\sum _{I=1} ^{M}
\left(O^1_{I+1} -2O^1_I+O^1_{I-1} \right)\omega^{nI}\cr
&=&\left[M+1+\frac{g_{YM}^2MN}{8\pi^2}2\left(1-\cos \frac{2\pi
n}{M}\right)\right]O_{n,1} \label{ho1i} \eea In the large $M$
limit we can expand the cosine up to ${\cal O}(1/M^2)$.  We also
recall that, in this case $2p^-=D-M$ to get \be 2p^-\circ
{\cal{O}}_{n,1} =\left[1+\frac{1}{2}\frac{g_{\rm
YM}^2N}{M}n^2\right]O_{n,1} \label{autov1} \ee The string theory
result is given in (\ref{1impspec}), expanding it in powers of
$g_s {N\over M}={ g_{YM}^2 N\over 4\pi M}$ one finds
that (\ref{autov1}) provides the exact one loop correction. This
confirms the proposed operator-state map to this order in
expansion in the world-sheet momentum, $n$.

Now, consider the case where $k=2$. We find that the two states
\begin{eqnarray}
O_+=\sum_{I=1}^M e^{2\pi i mI/M}\left\{  {\rm
TR}(A_1...\Phi_I...A_MA_1...A_M)+ {\rm
TR}(A_1 ...\Phi_I...A_M){\rm TR}(A_1...A_M) \right\} \\
 O_-=\sum_{I=1}^M e^{2\pi i mI/M}\left\{   {\rm TR}(A_1...\Phi_I...A_MA_1...A_M)- {\rm
TR}(A_1 ...\Phi_I...A_M){\rm TR}(A_1...A_M) \right\}
\end{eqnarray}
are exact eigenstates of the $D$ with eigenvalues
\begin{eqnarray}
D\circ O_+ = \left[2M+1+\frac{ g_{YM}^2M(N+
1)}{4\pi^2}(1-\cos\frac{2\pi
m}{M}) \right]O_+ \\
D\circ O_- = \left[2M+1+\frac{ g_{YM}^2M(N-
1)}{4\pi^2}(1-\cos\frac{2\pi m}{M}) \right] O_-
\end{eqnarray}
respectively.

These eigenstates of the dilatation operator  have an interesting
form.  They are a mixture of one-trace and two-trace operators
which we would normally associate with one-string and two-string
states and the mixing does not depend on the coupling constant, so
does not go away when the coupling constant is made small.

Further, note that, in the double scaling limit, these eigenvalues
are degenerate.  This means that string  loop corrections vanish
in this limit and the free string spectrum is exact.

What about the case $k=3$?  In this case there are a number of
degenerate states
\begin{eqnarray}
\Op_1=\sum_{I=1}^Me^{2\pi i mI/M}{\rm TR}\left( A_1...\Phi_I...A_M(A_1...A_M)^2\right) \\
\Op_2=\sum_{I=1}^Me^{2\pi i mI/M}{\rm TR}\left( A_1...\Phi_I...A_M(A_1...A_M) \right) {\rm TR}\left(  A_1...A_M\right) \\
\Op_3=\sum_{I=1}^Me^{2\pi i mI/M}{\rm TR}\left( A_1...\Phi_I...A_M\right){\rm TR}\left((A_1...A_M)^2\right) \\
\Op_4=\sum_{I=1}^Me^{2\pi i mI/M}{\rm TR}\left( A_1...\Phi_I...A_M
\right) {\rm TR}\left( A_1...A_M\right){\rm TR}\left(
A_1...A_M\right)
\end{eqnarray}
All operators have the same tree level dimension $3M+1$. Here,
$k=3$ and $m$ is the total wrapping number of each state. Note
that the world-sheet momentum in $\Op_1$ is $n=m/3$, in $\Op_2$ is
$n=m/2$ and in $\Op_3$ and $\Op_4$ it is $n=m$. It satisfies the
level matching condition $n=km$ for single string states in each
case if we interpret a single trace as a single trace state.
Again, eigenvectors of the dilatation operator are found by
choosing simple linear combinations,
\begin{eqnarray}
D\circ(\Op_1+\Op_2-\Op_3-\Op_4) =
 \left[3M+1+\frac{ g_{YM}^2M(N-
1)}{4\pi^2}(1-\cos\frac{2\pi m}{M}) \right]
(\Op_1+\Op_2-\Op_3-\Op_4)\nonumber
\\
D\circ(2\Op_1+2\Op_2+\Op_3+\Op_4)=  \left[3M+1+\frac{ g_{YM}^2M(N+
2)}{4\pi^2}(1-\cos\frac{2\pi m}{M})
\right](2\Op_1+2\Op_2+\Op_3+\Op_4)
\nonumber \\
D\circ(\Op_1-\Op_2+\Op_3-\Op_4)= \left[3M+1+\frac{ g_{YM}^2M(N+
1)}{4\pi^2}(1-\cos\frac{2\pi m}{M}) \right](\Op_1-\Op_2+\Op_3-\Op_4)\nonumber\\
D\circ(2\Op_1-2\Op_2-\Op_3+\Op_4)=\left[3M+1+\frac{ g_{YM}^2M(N-
2)}{4\pi^2}(1-\cos\frac{2\pi m}{M})
\right](2\Op_1-2\Op_2-\Op_3+\Op_4)\nonumber
\end{eqnarray}
 Again, we see that the degeneracy is split, but by
terms which will vanish in the double scaling limit.  In that
limit, the four states are degenerate again.

The situation will be similar with any value of $k$.  Generally,
the dilatation operator mixes operators with different
distributions of traces.   However, in the case of a single
impurity, is straightforward to show that this mixing vanishes
in the double scaling limit.  This gives the one-impurity state an
interesting property, they get Yang-Mills loop corrections to all
orders in perturbation theory from planar diagrams, but all
non-planar diagrams vanish.  This is interpreted in the string
theory as the absence of string loop corrections to the free
string spectrum.

\section{Two impurities, $k=1$ }

Let us now consider the string state with two oscillators and
one-unit of light cone momentum $k=1$. The gauge theory operators
dual to this state can be obtained for example by a double
insertion of $\Phi$ fields into a string of $A$ fields \be
O^1_{IJ}= {\rm TR}
(A_1 A_2\cdots A_{I-1} \Phi_I A_I \cdots A_{J-1} \Phi_J A_J \cdots A_{M})\\
\label{2impk=1} \ee This is a set of $M(M+1)/2$ independent
operators $(I\leq J)$.  At tree level, they are degenerate, with
conformal dimension $M+2$.

The action of the  dilatation operator (\ref{dilatationoperator})
on the states (\ref{2impk=1}) is given by 
\be 
(D-2M-2)\circ
O^1_{IJ}= \frac{g^2_{YM}MN}{8\pi^2
}\left(-\left(\nabla_I^2+\nabla^2_J\right)O^1_{IJ}
-\delta_{IJ}\left(\nabla_J-\hat{\nabla}_I\right)O^1_{IJ}\right)
\label{hamoij} 
\ee 
where  we have introduced the forward and
backwords shift operators defined by \bea &&\nabla_I
O^1_{IJ}=O^1_{I+1,J}-O^1_{IJ}\cr
&&\hat{\nabla}_IO^1_{IJ}=O^1_{IJ}-O^1_{I-1,J} \eea respectively.
The lattice laplacian with respect to each variable $I$ or $J$ is
defined as \bea \nabla_I^2 O^1_{IJ}=
O^1_{I+1,J}-2O^1_{IJ}+O^1_{I-1,J} \label{nabla} \eea The second
term on the right hand side of eq.(\ref{hamoij}) is a contact term
that originates when $I=J$.  The problem for finding the spectrum
of the operator in (\ref{hamoij}) is treated carefully in Appendix
B.  There, it is found that, in the large $M$ limit, the problem
of finding the spectrum of the difference operator with $I\leq J$
and the contact term can be replaced by simply looking for the
spectrum of the difference operator operating on symmetric, doubly
periodic functions and no contact term. For simplicity, we will
implement this procedure here.  To begin, we define a symmetric
function $$ O_{IJ}=\left\{ \matrix{ {\rm
TR}(A_1...\Phi_I...\Phi_J...A_M) & I\leq J \cr {\rm
TR}(A_1...\Phi_J...\Phi_I...A_M) & I>J \cr}\right. $$ Then, we
must take the large $M$ limit.

Introducing the continuous variables $x=I/M$ and $y=J/M$, and
taking $M$ large, $x,y $ take values between 0 to 1. Consequently,
the continuum limit of equation (\ref{hamoij}) reads \be \left(
2p^--2\right)\circ O^1(x,y)=\frac{g^2_{YM}N}{8\pi^2 M}\left( -
\left(\partial_x^2+\partial_y^2\right) -
\delta(x-y)\left(\partial_x-\partial_y\right)\right)O^1(x,y)
\label{hamoxy} \ee where   $\delta_{IJ}\to {\delta(x-y)\over M}$
in the continuum limit and $2p^-=D-M$. As anticipated, if we
assume that $O^1(x,y)$ is symmetric, the last term goes away in
the continuum limit. Now, taking into account the fact that
$O(x,y)$ is periodic in each variable with period 1, we find
 \be 2p^-\circ
{\cal{O}}_{n_1,n_2,1}=\left[2+\frac{1}{2}\frac{g_{YM}^2N}{M}(n_1^2+n_2^2)
\right]O_{n_1,n_2,1} \label{autov2} \ee This result reproduces the
requisite string energy spectrum (\ref{sspectrum}) for the case of
two oscillators up to one loop order. It shows that for $k=1$ the
operators corresponding to winding states with two oscillators are
free string states since they get only planar corrections to the
anomalous dimension. More generally this result holds
independently on the number and type of impurities one considers
since the single trace operator cannot be split into multi-trace
operators by the action of the effective interaction hamiltonian.

\section{Two impurities, $k=2$}

In this section we will study  gauge theory operators with two
impurity $\Phi$ fields that describe the string theory sector with
discrete light-cone  momentum $k=2$.   The basis of operators with
two impurities is \bea &&O^{C2}_{IJ}= {\rm TR} (A_1 A_2\cdots A_{I-1}
\Phi_I A_I \cdots A_{J-1} \Phi_J A_J \cdots A_{M}A_1\cdots
A_M)~~,~~~I\leq J\cr &&\cr &&O^{S2}_{IJ}= {\rm TR} (A_1 A_2\cdots
A_{I-1} \Phi_I A_I \cdots A_M A_1\cdots A_{J-1} \Phi_J A_J \cdots
A_{M})\cr &&\cr &&O^{C11}_{IJ}= {\rm TR} (A_1 A_2\cdots A_{I-1}
\Phi_I A_I \cdots A_{J-1} \Phi_J A_J \cdots A_{M}) {\rm TR}(A_1\cdots
A_M)~~,~~~I\leq J\cr &&\cr &&O^{1}_{I}O^{1}_{J}= {\rm TR} (A_1
A_2\cdots A_{I-1} \Phi_I A_I \cdots A_M) {\rm TR} (A_1 A_2\cdots
A_{J-1} \Phi_J A_J \cdots A_M) 
\label{2impk=2} 
\eea 
Again, there
are linear combinations of these operators which are special.  For
example,
\begin{equation} 
(D-2M-2)\circ\left( O^{C2}_{IJ}+
O^{S2}_{IJ}\right)=\frac{g_{YM}^2MN}{8\pi^2}\left(-\nabla_I^2-\nabla_J^2-\delta_{IJ}(\nabla_J
-\hat\nabla_I)\right)\left( O^{C2}_{IJ}+ O^{S2}_{IJ}\right)
\end{equation}
It is also periodic in that $$ O^{C2}_{IM+1}+ O^{S2}_{IM+1} =
O^{C2}_{1I}+ O^{S2}_{1I}
$$
We show in Appendix B that, in the double scaling limit, this
operator with this boundary condition has the spectrum
\begin{equation}
2p^-\circ\left( O^{C2}_{IJ}+ O^{S2}_{IJ}\right)=\left[
2+\frac{1}{2}\frac{g_{YM}^2N}{M}\left(n_1^2+n_2^2\right)\right]\left(
O^{C2}_{IJ}+ O^{S2}_{IJ}\right)
\end{equation}

In a similar way, we can see that the combination $ O^{C11}_{IJ}+
O^{1}_{I}O^1_J$ is mixed with the other operators, but the mixing
vanishes in the double scaling limit.  It then obeys the same
equation with the same boundary condition as $O^{C2}_{IJ}+
O^{S2}_{IJ}$ and therefore has the same spectrum

\begin{equation}
2p^-\circ\left( O^{C11}_{IJ}+ O^{1}_{I}O^1_J\right)=\left[
2+\frac{1}{2}\frac{g_{YM}^2N}{M}\left(n_1^2+n_2^2\right)\right]\left(
O^{C11}_{IJ}+ O^{1}_{I}O^1_J\right)
\end{equation}

These are half of the allowed states.  Recall that the
two-oscillator state of the string, expanded to the leading order
had spectrum
$$
2p^-= 2+\frac{1}{2}\frac{g_{YM}^2N}{M}\left(
\left(\frac{n_1}{2}\right)^2+\left(\frac{n_2}{2}\right)^2\right)+\ldots
~~,~~n_1+n_2=2\cdot{\rm integer}$$. The latter is the level
matching condition.  It implies that $n_1$ and $n_2$ are either
both even or both odd.  In the above, we have found two towers of
states where they are both even.  They could be associated
with two string states with $k=2$ and two oscillators excited,
both with even world-sheet momenta.

Finally, there are two states left. Operator mixing can be
diagonalized in the double scaling limit by taking the linear
combinations
\begin{eqnarray}
O^+_{IJ}=O^{C2}_{IJ}- O^{S2}_{IJ}+O^{C11}_{IJ}-O^{1}_{I}O^1_J
\\
O_{IJ}^-=O^{C2}_{IJ}- O^{S2}_{IJ}-O^{C11}_{IJ}+ O^{1}_{I}O^1_J
\end{eqnarray}
These have the boundary condition that $$ O^+_{IM+1}=-O^-_{1I}$$
In the double scaling limit, they obey the equations
\begin{eqnarray}
(D-2M-2)\circ
O^+(x,y)=\frac{g_{YM}^2MN}{8\pi^2}\left(-\partial_x^2-\partial_y^2
-2\frac{M}{N}\epsilon(x-y)(\partial_x-\partial_y)\right)O^+(x,y)\label{eqn1}\\
(D-2M-2)\circ
O^-(x,y)=\frac{g_{YM}^2MN}{8\pi^2}\left(-\partial_x^2-\partial_y^2
+2\frac{M}{N}\epsilon(x-y)(\partial_x-\partial_y)\right)O^-(x,y)\label{eqn2}
\end{eqnarray}
We comment that these operators were originally defined only for
$x\leq y$ and with a contact interaction.  We have used the trick
discussed in Appendix B of extending the function symmetrically to
all $x$ and $y$ so that $O^\pm(x,y)=O^{\pm}(y,x)$ and cancelling
the contact interaction. (We can check after we have found a
solution that the contact term operating on it indeed vanishes.)
The boundary conditions are then
$$
O^+(x+1,y)=O^+(x,y+1)=O^-(x,y)~~,~~O^-(x+1,y)=O^-(x,y+1)=O^+(x,y)
$$
These are compatible with the above equation if we extend the
antisymmetric step function so that
\bea
&&\epsilon(x)=1~,~ x\in (-2,1)~,~(0,1)~,~(2,3),...\cr
&&\epsilon(x)=-1~,~ x\in (-1,0)~,~(1,2)~,~(3,4),...
\label{step}
\eea
Then $\epsilon(x\pm 1)=-\epsilon(x)$.

It is natural to introduce the center of mass and the relative
coordinates $R$ and $r$ defined as \be
R=\frac{x+y}{2}~~,~~~~~r=y-x \label{relcoor} \ee where $0\leq
R\leq 1$ and $-1\leq r \leq 1$. Then, the variables separate and
we can make the ansatz $$O^{+}(R,r)=e^{2\pi i
mR}u(r)~~,~~O^-(R,r)=e^{2\pi i m R}v(r)$$where, since the
functions should be periodic in $R$, $m$ is an integer.

The eigenvalues of the dilatation operator are
$$
D=2M+2+\frac{g_{YM}^2N}{8\pi^2M}\left(\lambda+2\pi^2 m^2\right)
$$
where $\lambda$ are eigenvalues obtained by solving the equations

%\begin{figure}[h]
%\begin{center}
%\epsfig{file=marta1.eps,width=120mm}
%\end{center}
%\caption{The sign function $\epsilon(x-y)$ has to be considered a
%periodic function of period 2. }
%\end{figure}

\bea
&&\left(\partial^2_r-2g_2\epsilon(r)\partial_r+\frac{\lambda}{2}\right)u(r)=0\cr
&&\left(\partial^2_r+2g_2\epsilon(r)\partial_r+\frac{\lambda}{2}\right)v(r)=0
\label{eqautov} \eea   The equation should now be solved with the
boundary condition
\begin{equation}u(r+1)=-(-1)^mv(r)\label{bc1}\end{equation} This boundary condition
has already been used to set the eigenvalues equal in
(\ref{eqautov}).  We have also denoted
\be
g_2=\frac{M}{N}
\label{g2}
\ee

 Consider the equation for the function
$u(r)$ (the equation for $v(r)$ is identical with $g_2\to - g_2$)
\bea
\cases{\left(\partial^2_r-2g_2\partial_r+\frac{\lambda}{2}\right)u(r)=0~~~,~~~~~r>0\cr
\left(\partial^2_r+2g_2\partial_r+\frac{\lambda}{2}\right)u(r)=0~~~,~~~~~r<0}
\label{first} \eea The solution of this one dimensional eigenvalue
problem is trivial. The solution for positive $r$ can be written
as \be u(r)_+=ae^{\omega_+r}+be^{\omega_-r} \label{u+} \ee where
$\omega_{\pm}=g_2\pm\sqrt{g_2^2-\frac{\lambda}{2}}$. For negative
$r$ one has \be u(r)_-=ce^{\omega'_+r}+de^{\omega'_-r} \label{u-}
\ee where $\omega'_{\pm}=-g_2\pm\sqrt{g_2^2-\frac{\lambda}{2}}$.

By requiring the continuity of the function and its first
derivative in $r=0$ and the continuity of the function in $r=1$ we
find \bea u(r)=a\cases{
\left(\sqrt{g_2^2-\frac{\lambda}{2}}-g_2\right)e^{\left(g_2+\sqrt{g_2^2-\frac{\lambda}{2}}\right)r}
+\left(\sqrt{g_2^2-\frac{\lambda}{2}}+g_2\right)e^{\left(g_2-\sqrt{g_2^2-\frac{\lambda}{2}}\right)r}
& $r> 0$ \cr
\left(\sqrt{g_2^2-\frac{\lambda}{2}}+g_2\right)e^{\left(-g_2+\sqrt{g_2^2-\frac{\lambda}{2}}\right)r}
+\left(\sqrt{g_2^2-\frac{\lambda}{2}}-g_2\right)e^{-\left(g_2+\sqrt{g_2^2-\frac{\lambda}{2}}
\right)r} &$r<0$}\ . \label{solu} \eea where $a$ is a
normalization constant that we will keep undetermined. Then we
should impose the boundary conditions on the function and its
derivative. The function $u(r)$ and its derivative must be
periodic of period 2. Since $u(r)$ is symmetric $u(1)=u(-1)$, but
requiring that $u'(1)=u'(-1)$ implies that $u'(1)=0$. The latter
condition leads to 
\be
\frac{\lambda}{2}\left(e^{-\sqrt{g_2^2-\frac{\lambda}{2}}}-
e^{\sqrt{g_2^2-\frac{\lambda}{2}}}\right)=0 \label{sol} \ee which
determines the eigenvalue \bea \lambda=\cases{0\cr
2g_2^2+2\pi^2n^2} 
\label{lambdau} 
\eea 
the solution $\lambda=0$
has to be discarded because to it corresponds a constant
function $u(r)$ which is not compatible with the boundary condition (\ref{bc1}). 
Moreover $n$ has to be different from
zero. In fact, if we take $n$ to be zero, then $\lambda =2g_2^2$.
But this solution for $\lambda$ implies $u(r)=0$ as can be easily
seen from equation (\ref{solu}). Since the eigenvalue depends only
on $g_2^2$ the same eigenvalue will be found for the solution
$v(r)$. Finally, the light-cone momenta for these two states are

$$
2p^-=2+\frac{g_{YM}^2N}{8\pi^2M}\left(2\pi^2n^2+2\pi^2
m^2+2g_2^2\right)
$$
Now, if we take $m=\frac{1}{2}(n_1+n_2)$ and
$n=\frac{1}{2}(n_1-n_2)$ where $n_1\pm n_2$ are necessarily even,
we get
$$
2p^-=2+\frac{1}{2}\frac{g_{YM}^2N}{ M}\left( n_1^2+n_2^2
 \right)+\frac{g_{YN}^2M}{4\pi^2 N}
$$
We must still impose the boundary condition (\ref{bc1}).  To get
$v(r)$, we change the sign of $g_2$ in $u(r)$.  Then, we see that
there are two possibilities for the integers $m$ and $n$: either
$m$ is even and $n$ is odd, or $m$ is odd and $n$ is even.  In
either case, this implies that both $n_1$ and $n_2$ are odd
integers.

Let us review:  Level matching, $n_1+n_2=2\cdot$integer, requires
that the world-sheet momenta $n_1$ and $n_2$ are either both even
integers or both odd integers.  When they are both even, the free
string spectra are not corrected by string loops, at least to the
leading order in world-sheet momenta.  When they are both odd,
they are corrected by string loops. One way to present the
correction is to rename the constant which governs the world-sheet
energy
$$
\tilde \alpha =\frac{g_{YM}^2N}{ M}
$$
Then we can write the above formula as
\begin{equation}
2p^-=2+\frac{1}{2}\tilde\alpha \left( n_1^2+n_2^2
 \right)+\frac{4}{\tilde\alpha}g_s^2
\label{exactspeck2}\end{equation} This is an exact result for
string states with two units of light-cone momentum, $k=2$. As
expected, the first two terms on the right-hand-side of
(\ref{exactspeck2}) give the free string spectrum.  What is
surprising is that the interaction term truncates at second order
in the closed string coupling.  In principle, the eigenvalue could
have been a complicated function of $g_2=N/M= 4\pi
g_s/\tilde\alpha$ and could have generated all orders in the
string coupling.  We do not presently have an understanding of why
it should truncate at second order.  This truncation is a definite
prediction for string loop corrections which should be checked in
string theory.

\section{Two impurities, k=3}

In this section we will consider the gauge theory operators with two
insertion of $\Phi$ fields
that describe the string theory sector with DLCQ momentum $k=3$. 
The basis of operators with two impurities 
is made of the following 9 operators
\be
O^{S3}_{IJ},~~O^{S3}_{JI},~~O^{C3}_{IJ},~~O^{C21}_{IJ},~~O^{S21}_{IJ},~~O^{C12}_{IJ},
~~O^{C111}_{IJ},
~~O^{2}_{I}O^{1}_{J},~~O^{1}_{I}O^{11}_{J}
\label{basis}
\ee
where
\bea
&&O^{C3}_{IJ}= {\rm TR}
[A_1 A_2\cdots A_{I-1} \Phi_I A_I \cdots A_{J-1} \Phi_J A_J \cdots
A_{M}(A_1\cdots A_M)^2]~~
,~~I\leq J\cr
&&O^{S3}_{IJ}= {\rm TR}
[A_1 A_2\cdots A_{I-1} \Phi_I A_I \cdots A_M A_1\cdots A_{J-1} \Phi_J A_J
\cdots A_{M}A_1\cdots A_M]\cr
&&O^{C21}_{IJ}= {\rm TR}
[A_1 A_2\cdots A_{I-1} \Phi_I A_I \cdots A_{J-1} \Phi_J A_J \cdots
A_{M}A_1\cdots A_M]
{\rm TR}[A_1\cdots A_M]~~,~~I\leq J\cr
&&O^{S21}_{IJ}= {\rm TR}
[A_1 A_2\cdots A_{I-1} \Phi_I A_I \cdots A_{M}A_1\cdots A_{J-1} \Phi_J
A_J \cdots A_{M}]
{\rm TR}[A_1\cdots A_M]~~,~~I\leq J\cr
&&O^{C12}_{IJ}= {\rm TR}
[A_1 A_2\cdots A_{I-1} \Phi_I A_I \cdots A_{J-1} \Phi_J A_J \cdots A_{M}]
{\rm TR}[(A_1\cdots A_M)^2]~~,~~I\leq J\cr
&&O^{C111}_{IJ}= {\rm TR}
(A_1 A_2\cdots A_{I-1} \Phi_I A_I \cdots A_{J-1} \Phi_J A_J \cdots A_{M})
\left({\rm TR}[A_1\cdots A_M]\right)^2~~,~~I\leq J \cr
&&O^{2}_{I}O^{1}_{J}= {\rm TR}
[A_1 A_2\cdots A_{I-1} \Phi_I A_I \cdots A_MA_1\cdots A_M]
{\rm TR} [A_1 A_2\cdots A_{J-1} \Phi_J A_J \cdots A_M]\cr
&&O^{1}_{I}O^{11}_{J}= {\rm TR}
[A_1 A_2\cdots A_{I-1} \Phi_I A_I \cdots A_M]
{\rm TR} [A_1 A_2\cdots A_{J-1} \Phi_J A_J \cdots A_M]{\rm TR}[A_1\cdots A_M]\cr
&&
\label{2impk=3}
\eea
As explained in Appendix B, in the large $M$ limit, the extension to all $I$ and $J$
of the functions $O_{IJ}$ 
defined only for $I\leq J$, becomes a symmetric doubly periodic function.
There are linear combinations of these states for which 
the action of $(D-3M-2)$
reduces to that of the Laplacian. Two states of this type which are periodic
of period 1 are
\bea
&&u_{IJ}^1=O_{IJ}^{C21}+O_{IJ}^{S21}+\frac{1}{2}\left(O_I^1O_J^2+
O_I^{2}O_J^1\right)\cr
&&u_{IJ}^2=\frac{1}{2}O_{IJ}^{C111}+O_I^1O_J^{11}
\label{lapleigen}
\eea
These state are periodic in that
\be
u_{IM+1}^1=u_{1I}^1~~~,~~~u_{IM+1}^2=u_{1I}^2
\label{bound}
\ee
In the double scaling limit these operators that satisfy the boundary condition
(\ref{bound}), have the spectrum
$$
2p^-\circ u^i(x,y)=(2-\frac{g^2_{YM}N}{8\pi ^2M}
\nabla ^2) u^i(x,y)=(2+\frac{g^2_{YM}N}{8\pi ^2M}\lambda )u_{IJ}^i
$$ 
To get a
string state corresponding
to 3 units of light-cone momentum, the solution to this equation has to
be put into the
form $\exp[{2 i \pi (n_1 x+n_2 y)/ 3}]$ as in (\ref{excstatek}). As in
the $k=2$ case however,
since these states must be periodic of period 1 both in $x$ and $y$, $n_1$ and $n_2$ must be
multiples of 3.
 $\lambda$ then is given by
$\lambda=4\pi ^2(n_1^2+n_2^2)/9$, providing an anomalous dimension
for these operators $\Delta=g_{YM}^2N(n_1^2+n_2^2)/(18 M)$.
This is again an exact result, the string states corresponding to
these operators are free, the one-loop anomalous dimension
provides in fact the expansion to order $g_{YM}^2$ of the free
string spectrum. $\Delta$ does not get corrections beyond the
planar level namely from string interactions. However, only string
states created by oscillators with $n_1$ and $n_2$ multiples of 3
behave as free string states.

We are left with 7 independent states, which can be reorganized in a
more convenient way
by introducing the following combinations of double and triple trace
operators
\bea
&&D^{12}_{IJ}=O^{C12}_{IJ}-O_I^{1}O^2_J-O_I^{2}O^1_J\cr
&&T_{IJ}^{111}=O_{IJ}^{C111}-O_I^1O_J^{11}
\label{newdef}
\eea
These combinations are orthogonal to (\ref{lapleigen}) and together with
the operators $O^{S3}_{IJ}$, $O^{S3}_{JI}$, $O^{C3}_{IJ}$, $O^{C21}_{IJ}$,
$O^{S21}_{IJ}$ form a closed basis for $(D-3M-2)$.
Consider a general continuous symmetric linear combination of these states
and call it $u_{IJ}$
\bea
u_{IJ}&=&\theta (J-I)\left[c_1O^{S3}_{IJ}+c_2O^{S3}_{JI}+
c_3O^{C3}_{IJ}+c_4O^{C21}_{IJ}+c_5O^{S21}_{IJ}+c_6D^{12}_{IJ}+c_7T^{111}_{IJ}\right]\cr
&+&
\theta (I-J)\left[c_2 O^{S3}_{IJ}+c_1 O^{S3}_{JI}+
c_3O^{C3}_{IJ}+c_4O^{C21}_{IJ}+c_5O^{S21}_{IJ}+c_6D^{12}_{IJ}+c_7T^{111}_{IJ}\right]
\label{u}
\eea
where we have introduced the Heaviside theta function defined as
$$
\theta(J-I)=\cases{1~~~~~J>I\cr
1/2~~~J=I\cr
0~~~~J<I}
$$
Acting with $(D-3M-2)$ on $u_{IJ}$ and taking the double scaling limit, 
one discovers that the system of equations
closes after three
iteration of the action of $(D-3M-2)$, namely
\bea
\cases{
(D-3M-2)\circ u(x,Y)=\frac{g^2_{YM}N}{8\pi^2M}\left(-\nabla^2u(x,y)-\frac{M}{N}
\epsilon(x-y)\left(\partial_x-\partial_y\right)w(x,y)\right)\cr
(D-3M-2)\circ w(x,y)=\frac{g^2_{YM}N}{8\pi^2M}\left(-\nabla^2w(x,y)-\frac{M}{N}
\epsilon(x-y)\left(\partial_x-\partial_y\right)v(x,y)\right)\cr
(D-3M-2)\circ v(x,y)=\frac{g^2_{YM}N}{8\pi^2M}\left(-\nabla^2v(x,y)-9\frac{M}{N}
\epsilon(x-y)\left(\partial_x-\partial_y\right)w(x,y)\right)}
\label{set1}
\eea
where $w(x,y)$  and $v(x,y)$ are written in terms of the coefficients
of $u(x,y)$ in Appendix C and the step function $\epsilon(x-y)$ 
has to be periodically continued as in (\ref{step}).
It is not difficult now to decouple this system of equations. There is
in fact
a solution to which correspond free string states. It is given by the solutions of the
equations
\be
w(x,y)=v(x,y)=0,~~~~~~~~~~~~~~~~(D-3M-2)\circ u(x,y) =-\frac{g^2_{YM}N}{8\pi^2M}
\nabla^2 u(x,y)
\label{free3k}
\ee
The first two equations in (\ref{free3k}) fix the values of the coefficients 
$c_i$ to give for $u(x,y)$
three independent solutions corresponding to three states periodic of
period 1 both in $x$ and $y$ 
\bea
&&u^3(x,y)=O^{S3}(x,y)+O^{S3}(y,x)+O^{C3}(x,y)\cr
&&u^4(x,y)=\theta(y-x)[O^{S3}(y,x)+O^{C3}(x,y)-\frac{1}{3} T^{111}(x,y)]\cr
&&+\theta(x-y)[O^{S3}(x,y)+O^{C3}(x,y)-\frac{1}{3} T^{111}(x,y)]\cr
&&u^5(x,y)=O^{C21}(x,y)+O^{S21}(x,y)-\frac{1}{4}D^{12}(x,y)
\label{lapleigen1}
\eea
These states have the same free string spectrum of $u^1$ and $u^2$.
Since they also have
the same periodicity one has to require the same condition on the
oscillators generating
these string states, namely that the levels to which they correspond
($n_1$ and $n_2$) must be multiples of 3.

We have still to find four states.
Taking the linear combinations
\be
\psi_{\pm}(x,y)=3w(x,y)\pm v(x,y)
\label{psipm}
\ee
from (\ref{set1}), one gets the equations
\bea
\cases{
(D-3M-2)\circ \psi_{+}(x,y)=\frac{g^2_{YM}N}{8\pi^2M}\left(-\nabla^2\psi_+(x,y)
-3\frac{M}{N}\epsilon(x-y)
\left(\partial_x-\partial_y\right)\psi_+(x,y)\right)\cr
(D-3M-2)\circ \psi_{-}(x,y)=\frac{g^2_{YM}N}{8\pi^2M}\left(-\nabla^2\psi_-(x,y)
+3\frac{M}{N}\epsilon(x-y)
\left(\partial_x-\partial_y\right)\psi_-(x,y)\right)\cr
(D-3M-2)u(x,y)=\frac{g^2_{YM}N}{8\pi^2M}\left(-\nabla^2u(x,y)-\frac{M}{6N}\epsilon(x-y)
\left(\partial_x-\partial_y\right)(\psi_{+}+\psi_{-})\right)}\cr
\label{set3}
\eea
where the first two equations are decoupled. In Appendix C we show that
in both
$\psi_\pm$ there are only two independent coefficients of the original 9 basis
operators
(\ref{basis}).
Therefore $\psi_\pm$
provide the remaining 4 states. As we will see
in the next section these states
correspond to operators that are periodic of period 6 and get computable
corrections to all orders in the genus expansion.

\subsection{Solution}

In this subsection we solve the equations (\ref{set3}) for $\psi_{\pm}(x,y)$.
%\begin{figure}[h!]
%\begin{center}
%\epsfig{file=marta.eps,width=120mm}
%\end{center}
%\caption{The periodic form of the sign function in the region of
%periodicity of
%the function $\psi_{\pm}$.}
%\label{epsilon}
%\end{figure}
%\figuren{marta}{120mm}{}
To this purpose, it is convenient to introduce again the center of mass
and relative coordinates (\ref{relcoor}). Let us focus on $\psi_+(R,r)$.
The variables separate and we can write
$$
\psi_+(R,r)=e^{2 i \pi m R}\psi_+(r)
$$
where $m$ is an integer.
The eigenvalues of the dilatation operator are
\be
D=3M+2+\frac{g^2_{YM}N}{8\pi^2M}\left(\lambda +2\pi^2m^2\right)
\ee
The eigenvalues $\lambda$ can be obtained by solving the equation
\be
\left(\partial^2_r-3g_2\epsilon(r)\partial_r+\frac{\lambda}{2}\right)\psi_+(r)=0
\label{eqautovpsi+}
\ee
where $g_2$ is defined in (\ref{g2}).
The states we are considering are in general periodic of period 6 namely
when
$r\to r+6$ (corresponding to $J\to J+6M$) $\psi_+$ goes to itself
$\psi_+(r+6)=\psi_+(r)$. The relative coordinate has a range $-1\le r\le1$
but it can be periodically continued to the range $-3\le r\le 3$ in order
to realize the correct periodicity for the functions $\psi_\pm(r)$.

We divide the interval between $r=-3$ and $r=3$ in six regions
and we impose that the solution of the equation (\ref{eqautovpsi+}) matches
at the boundary of each region.
We also can consider only half of the total
region, namely the region $0\leq r\leq 3$, because the function
$\psi_+(r)$ is symmetric for $r\to -r$.

In the regions I, $0\leq r\leq 1$, 
the solution of the equation (\ref{eqautovpsi+})  can be written as
\be
\psi_+(r)=
a\left(\omega_-e^{\omega_+r}-\omega_+e^{\omega_-r}\right)
\label{psi12}
\ee
where
$\omega_{\pm}=\frac{3}{2}g_2\pm\sqrt{\frac{9}{4}g_2^2-\frac{\lambda}{2}}$.
In the region II, $1\leq r\leq 2$, we write the solution as
\be
\psi_+(r)=be^{\eta_+r}+ce^{\eta_-r}
\label{4}
\ee
where
$\eta_{\pm}=-\frac{3}{2}g_2\pm\sqrt{\frac{9}{4}g_2^2-\frac{\lambda}{2}}$.
For $r=1$ we have to require the continuity of the function
\be
a\left(\omega_-e^{\omega_+}-\omega_+e^{\omega_-}\right)
=be^{\eta_+}+ce^{\eta_-}
\label{match}
\ee
so that, in the region II, we get 
\be
\psi_+(r)=\frac{ae^{\eta_+(r-2)}}{\sqrt{9g_2^2-2\lambda}}
\left[\lambda e^{\sqrt{9g_2^2-2\lambda}}-3g_2\omega_+\right]+
\frac{ae^{\eta_-(r-2)}}{\sqrt{9g_2^2-2\lambda}}
\left[\lambda e^{-\sqrt{9g_2^2-2\lambda}}-3g_2\omega_-\right]
\label{cb}
\ee
Finally, in the region III, $2\leq r\leq 3$, we write
\be
\psi_+(r)=de^{\omega_+r}+fe^{\omega_-r}
\ee
Requiring again the continuity of the function and of its derivative in
$r=2$,
we have that in the region III,
\bea
\psi_+(r)&=&\frac{ae^{\omega_+(r-2)}}{(9g_2^2-2\lambda)}
\left[9g_2^2\omega_-+3g_2\lambda\left(1-e^{-\sqrt{9g_2^2-2\lambda}}\right)-
2\lambda \omega_-e^{\sqrt{9g_2^2-2\lambda}}\right]\cr
&-&\frac{ae^{\omega_-(r-2)}}{(9g_2^2-2\lambda)}
\left[9g_2^2\omega_++3\lambda g_2\left(1-e^{\sqrt{9g_2^2-2\lambda}}\right)-
2\lambda \omega_+e^{-\sqrt{9g_2^2-2\lambda}}\right]
\label{psi5}
\eea
The function $\psi_+(r)$, by construction, is periodic of period 6 and is symmetric for $r\to -r$.
By requiring that $\psi'_+(3)=\psi'_+(0)=0$, we get an exact transcendental
equation
for $\lambda$
\bea
\frac{\lambda e^{-3(-g_2+\sqrt{9g_2^2-2\lambda})/2}}{18g_2^2-4\lambda}
\left(1-e^{\sqrt{9g_2^2-2\lambda}}\right)\left(2\lambda(1+e^{2\sqrt{9g_2^2-2\lambda}})
+e^{\sqrt{9g_2^2-2\lambda}}(2\lambda -27g_2^2)\right)=0\cr
\label{fine}
\eea
The differential equation for $\psi_-$ can be obtained
from the one for $\psi_+$
by changing the sign of $g_2$. Moreover, the eigenvalue equation, as it
should,
depends only on $g_2^2$. Consequently, one gets an identical transcendental
relation for the eigenvalue
$\lambda_-$ by solving the equation for $\psi_-$.

The values of $\lambda$ that solve this equation are
$\lambda =0$, $\lambda =2\pi ^2n^2+\frac{9}{2}g_2^2$ or the solutions of
\be
f(g_2^2,\lambda)=4\lambda \cosh {\sqrt{9g_2^2-2\lambda}}+2\lambda -27g_2^2=0
\label{lamb}
\ee
$\lambda=0$ has to be discarded because to it corresponds a trivial constant solution.
$\lambda =2\pi ^2 n^2+\frac{9}{2}g_2^2$ corresponds to states with
period 2 in the variable $r$.
These states cannot be realized with linear combinations of the form
$\psi_\pm$
which are necessarily periodic of period 6. Thus if we require the
periodicity
allowed for the operators of the form $\psi_\pm$ one has to consider
only the
solutions of (\ref{lamb}).

{}From (\ref{lamb}) one could in principle compute the energy
eigenvalues of the gauge theory
to all orders in $g_2^2$ and verify that they are in agreement with
those obtained from
the dual string energy spectrum.

To obtain a genus expansion solution of the string spectrum one can
write the
eigenvalue $\lambda$ as a Taylor series
in powers of $g_2^2$ around $g_2=0$
\be
\lambda=\lambda_0+\frac{d\lambda}{dg_2^2}\bigg|_{\lambda_0,g_2=0}g_2^2
+\frac{1}{2}\frac{d^2\lambda}{(dg_2^2)^2}\bigg|_{\lambda_0,g_2=0}g_2^4+\frac{1}{3!}
\frac{d^3\lambda}{(dg_2^2)^3}\bigg|_{\lambda_0,g_2=0}g_2^6+\ldots
\label{series}
\ee
where $\lambda_0$ is the eigenvalue for $g_2=0$. {}From
(\ref{lamb}) with $g_2=0$ one finds that  $\lambda_0=2\pi^2n^2/9$ with
$n$ any integer different from zero which is not multiple of 3.

The derivatives of $\lambda$ as a function of $g_2^2$
can be obtained by taking the derivatives
of the implicit function (\ref{lamb}) and computing them at
$\lambda=\lambda_0$ and
$g_2=0$. In fact since $f(g_2^2,\lambda)=0$ also its differential must
vanish
\be
\frac{\partial f}{\partial g_2^2}dg_2^2+\frac{\partial f}{\partial
\lambda}d\lambda=0
\ee
{}From this equation one finds
\be
\frac{d\lambda}{dg_2^2}=-\frac{\partial f}{\partial g_2^2}
\left(\frac{\partial f}{\partial \lambda}\right)^{-1}
\ee
and from the last result we can compute all the coefficients of the
equation (\ref{series}).

To the sixth order in $g_2$ the eigenvalue is 
\bea 
\lambda
&=&\frac{2}{9}\pi^2n^2+\frac{9\left(2\pi n\sin \frac{2\pi
n}{3}-9\right)} {2\left[2\pi n\sin \frac{2\pi n}{3}-9+12\left(\sin
\frac{\pi n}{3}\right)^2\right]}g_2^2\cr &+&\frac{2187(\sin
\frac{\pi n}{3})^3\left(6\pi n \sin \frac{\pi n}{3} -33\cos
\frac{\pi n}{3}-3(-1)^n\right)} {2\pi n\left(2\pi n\sin \frac{2\pi
n}{3}-3-6\cos \frac{2\pi n}{3}\right)^3}g_2^4\cr
&+&\left\{-2484\pi n -80\pi^3n^3+ 405\left(\sin \frac{2\pi
n}{3}+\sin \frac{4\pi n}{3}\right)\right.\cr &&\left. + 2\pi
n\left[(-243+4\pi^2 n^2)5\cos \frac{2\pi n}{3}
+27\right.\right.\cr &&\left.\left. +\pi n\left(4\pi n+16\pi n\cos
\frac{4\pi n}{3}+3(173\sin \frac{2\pi n}{3} +38\sin \frac{4\pi
n}{3})\right) \right]\right\}\cdot \cr &&\cdot
\frac{59049\left(\sin \frac{\pi n}{3}\right)^4} {8\pi^3
n^3\left(2\pi n\sin \frac{2\pi n}{3}-3-6\cos \frac{2\pi
n}{3}\right)^5}g_2^6 +\ldots 
\label{lambda4} 
\eea 
where $n$ is
related to the world-sheet momenta of the string excitations by
$n=n_1-n_2$. We remind here that $n$ has to be different from zero
and not a multiple of 3 otherwise $\lambda_0$ would not be a solution
of the transcendental equation (\ref{lamb}) for $g_2=0$. 
If it was a multiple of 3, $n=3l$, first
of all the solution (\ref{lambda4}) would collapse into the one
previously analyzed $\lambda=2\pi^2 l+\frac{9}{2}g_2^2$ which does
not have the correct periodicity. Moreover, in this case, since
$n_1+n_2=3m$ by the level matching condition, both $n_1$ and $n_2$
would be multiples of 3. For these string states however, we have
already shown that the predictions of the gauge theory are that
there are no string loop corrections to the free string spectrum.
Summarizing, when the world-sheet momenta $n_1$ and $n_2$ of the
string excitations are both multiples of 3, the gauge theory
predicts a free string spectrum. When they are not multiples of 3,
the anomalous dimension of the corresponding gauge theory
operators gets corrections to all non planar orders. For example
the spectrum of the two-oscillator state of the string, expanded up to
genus three is given by 
\be 
2p^-=2+\frac{g_{YM}^2N}{8 \pi^2
M}\left( 2\pi^2m^2+\lambda++O(g_2^8) \right)
\label{pmk3}
\ee 
To match the string spectrum in this case, here $m=(n_1+n_2)/3$ 
(this establishes the level matching condition) 
and $\lambda$ up to the sixth order in $g_2$ is given in
(\ref{lambda4}) with $n=n_1-n_2$. The procedure can be iterated to
obtain also higher order corrections to this spectrum.

\section{Summary and Conclusions}

In this paper we have studied non planar corrections to the spectrum of operators in the
${\mathcal N}=2$ supersymmetric Yang Mills theory which are dual to
string states in the maximally supersymmetric pp-wave background with a compact 
light-cone direction. The existence of a discrete light-cone momentum
simplifies the calculations of the anomalous dimension of gauge 
theory operators dual to string states.

The gauge theory predictions in the double scaling limit of large $N$ and long operators $M\to\infty$,
with ${M\over N}$ fixed, are:
\begin{itemize}

\item
String states with one oscillator and any value of the light cone momentum $k$ 
are quasi-protected in that they have an anomalous dimension which 
does not have corrections beyond the planar level, their string spectrum is free.
This is a correct prediction until $k$ is 
of order $N$ (or $M$), which is when operator mixing  sets in.

\item
String states with one unit of light cone momentum and any number of oscillators
are free states in that they do not get string loop corrections.

\item
For string states with two units of light cone momentum and two impurities
there are two possibilities: states for which both the world-sheet momenta are integer 
multiples of $k=2$, namely are even, have a free spectrum; states for which both the world-sheet momenta are
odd get only the one string loop correction given in (\ref{exactspeck2}).
The states with even-odd world-sheet momenta are excluded by level matching.

\item
String states with three units of light cone momentum and two impurities
for which both the world-sheet momenta 
are integer multiples of $k=3$, have a free spectrum. $k=3$ states for which both the world-sheet momenta are
not integer multiples of 3 get computable corrections to all orders, as in (\ref{pmk3}), 
where $\lambda$ is a solution of the exact equation (\ref{lamb}).
Up to three loop in the effective genus counting parameter expansion, $\lambda$ 
is given in (\ref{lambda4}).
\end{itemize}

The AdS/CFT correspondence should be checked wherever possible. Recently it were found
some discrepancies in the energy spectrum of BMN states with two string excitations,
computed in the framework of light-cone string field theory~\cite{Gutjahr:2004dv},
and that computed from the gauge theory~\cite{Beisert:2002bb,Beisert:2002ff,Beisert:2003tq}.
Even if such discrepancies might be originated by the choice of the cubic vertex of IIB strings
in a pp-wave background~\cite{Dobashi:2004nm,Lee:2004cq}, it would be extremely important
to check these results also in other contexts.
The
DLCQ of the string greatly simplifies the setting where the duality is realized, therefore
it might help in solving, one way or the other, for such discrepancies.
The calculations presented in this paper are precisely for the case of two string excitations
were the discrepancies were found in the usual pp-wave correspondence.

\acknowledgments

G.G. thanks A. Lerda for an interesting discussion, G.G. and M.O. are grateful to 
S. Mukhi for useful suggestions.

\appendix{\section{Anomalous dimensions}}

In this Appendix, we will compute the divergent parts which appear
in the one-loop corrections to the two-point function of composite
operators made from products of scalar fields. These loop
corrections arise from the Wick contractions of the operators in
the interaction Lagrangian with the operators inside the
composites.   Here, we will assume that the interaction Lagrangian
is quartic and has no derivatives.  We will also assume that there
are no derivatives in the composite operators.   Finally, we
assume that there are only  two contractions between the
interaction Lagrangian and each composite operator.   Then, the
one-loop correction to the two-point function is \be - \left<
\Op_\alpha (y) {\cal L}_F(x) \bar\Op_{\bar\beta} (0) \right>=
-\left< \Op_\alpha H \bar\Op_{\bar\beta}\right>_{MM}
[\Delta(y)]^{\frac{\Delta^0_\alpha+\Delta^0_\beta}{2}}~\int d^4x
\frac{\Delta^2(y-x)\Delta^2(x)}{\Delta^2(y)} \ee where $\Delta(y)$
is the scalar propagator. The product $\Delta^2(y-x)$ comes from
the two contractions between the interaction and $\Op_\alpha$ and
$\Delta^2(x)$ from contractions with $\bar \Op_{\bar\beta}$.  The
combinatorics of how the contraction is done are summarized in the
reduced matrix element $\left< \Op_\alpha H
\bar\Op_{\bar\beta}\right>_{MM}$. We shall use dimensional
regularization and work in $2\omega$-dimensions where the scalar
propagator is
$$
\Delta(x)=\frac{ \Gamma(\omega-1)}
{4\pi^\omega}\frac{1}{[x^2]^{\omega-1}}
$$
 The  divergent  one-loop correction has
the form \begin{eqnarray}&&\mu^{4-2\omega} \int d^{2\omega}x\frac{
\Delta^2(y-x)\Delta^2(x)}{\Delta^2(y)}\nonumber
\\&&\mu^{4-2\omega} \left(\frac{
\Gamma(\omega-1)}{4\pi^\omega}\right)^2
\frac{\Gamma(4\omega-4)}{\Gamma^2(2\omega-2)}\int_0^1d\alpha
\alpha^{2\omega-3}(1-\alpha)^{2\omega-3} \int d^{2\omega}x
\frac{[y^2]^{2\omega-2}}{[ x^2+y^2\alpha(1-\alpha)]^{4\omega-4}}
\nonumber \\&=&\mu^{4-2\omega} \left(\frac{
\Gamma(\omega-1)}{4\pi^\omega}\right)^2
\frac{\Gamma(4\omega-4)}{\Gamma^2(2\omega-2)}\int_0^1d\alpha
\alpha^{2\omega-3}(1-\alpha)^{2\omega-3}
\pi^\omega\frac{\Gamma(3\omega-4)}{\Gamma(4\omega-4)}\frac{[y^2]^{2\omega-2}}
{[y^2\alpha(1-\alpha)]^{3\omega-4}}
\nonumber \\ &=&\frac{1}{[\mu^2y^2]^{\omega-2}}
\frac{\Gamma^2(\omega-1)\Gamma(3\omega-4)}{16\pi^\omega\Gamma^2(2\omega-2)}
\frac{\Gamma^2(2-\omega)}{\Gamma(4-2\omega)}~=~\left(\frac{1}{8\pi^2(2-\omega)}
+\frac{\ln(\mu^2y^2)}{8\pi^2}+\ldots
\right)\label{compute}\end{eqnarray} where $\mu$ is the
renormalization scale.  The first, divergent term in the final
expression must be subtracted using a counterterm.  The remaining
logarithmic term shifts the exponent in the space-dependence of
the two-point function and thus contributes to the anomalous
dimension.  The terms denoted by dots are finite and
$y$-independent.

To do the above integrals, we have used the Feynman parameter
formula
$$
\frac{1}{A_1^{n_1}A_2^{n_2}}=\frac{\Gamma(n_1+n_2)}{\Gamma(n_1)\Gamma(n_2)}\int_0^1
d\alpha \alpha^{n_1-1}(1-\alpha)^{n_2-1}
$$
and the integral formulae
$$
\int d^{2\omega}x[x^2+m^2]^{-s}=\pi^\omega
\frac{\Gamma(s-\omega)}{\Gamma(s)}[m^2]^{\omega-s}
$$

$$
\int_0^1 d\alpha \alpha^{\mu-1}(1-\alpha)^{\nu-1}=\frac{
\Gamma(\mu) \Gamma(\nu) }{\Gamma(\mu+\nu) }
$$

The reduced matrix element is found by taking the most singular
terms in the operator product expansion
\begin{equation}
\left< \Op_\alpha H
\bar\Op_\beta\right>=D_\alpha^{~\gamma}\left<\Op_\gamma
\bar\Op_\beta\right>$$ where $$ ~:\Op_\alpha(y):~:{\cal L}_F(x):
=\Delta^2(x-y)D_\alpha^{~\beta}\Op_\beta(y)+{\rm less~singular}
\label{opexp}\end{equation}

 To one-loop order, $D_\alpha^{~\beta}$ is the dilatation
 operator.  Its eigenvalues are the one-loop contribution to the
 conformal dimensions of operators and its eigen-vectors are the operators.

\section{Contact terms in the planar limit}

In this Appendix we will consider the effect of contact terms
which appear in the leading order, planar interactions. We will
show that the contact terms do not vanish in the continuum limit.
However, they can be taken into account the particular symmetric
prescription for two-impurity operators which is used in this paper.

Here, we consider the composite operator
$$
{\cal O}_{IJ}(x)={\rm TR}\left( A_1(x)\ldots A_{I-1}(x)\Phi_I(x)A_I(x)\ldots
A_{J-1}(x)\Phi_J(x)A_J(x)\ldots A_M(x)\right)
$$
Hereafter, we shall drop the coordinate dependence, as it should
always be clear from the context that the composite operators in
question are defined by products of fields at the same point.  The
Hamiltonian acts on this operator as
\begin{eqnarray}
H_0{\cal O}_{IJ}=
\frac{g_{YM}^2MN}{8\pi^2}\left[{\cal O}_{I+1J}+
{\cal O}_{I-1J}-2{\cal O}_{IJ}+{\cal O}_{IJ+1}+{\cal O}_{IJ-1}-2{\cal
O}_{IJ}\right]
 ~\label{I<J}\\   \hfill ~1\leq I<J\leq &M&\nonumber\\
H_0
{\cal O}_{II}=\frac{g_{YM}^2MN}{8\pi^2}\left[{\cal O}_{I-1I}+
{\cal O}_{II+1}-2{\cal O}_{II}\right]~~,~~I=&J&
\label{I=J}\end{eqnarray} It is augmented by the boundary
condition
\begin{equation} {\cal O}_{IM+1}={\cal O}_{1I}\label{BC}\end{equation}

This resembles the problem of two free particles moving on a
latticized circle with a contact interaction.  The contact term is
important and does not vanish in the continuum limit, so must be
treated with some care.

 We seek an eigenstate of the operator defined by (\ref{I<J}),(\ref{I=J})
 and (\ref{BC}).  The operator in (\ref{I<J}) is just the lattice
 Laplacian which has eigenstate
\begin{equation}
\psi_{IJ}(p,\ell)=e^{ipI+i\ell J}+S(p,\ell) e^{i\ell I+ipJ}
\label{eigenstate}\end{equation} and eigenvalue
\begin{equation}
E=e^{ip}+e^{-ip}-2 + e^{i\ell}+e^{-i\ell}-2
\label{eigenvalue}\end{equation}

Here, we have noted that the eigenvalue problem is formally
symmetric under interchanging $I$ and $J$, so the eigenvalues
should carry a (perhaps projective) representation of the
permutation group. This representation is characterized by the
2-body S-matrix, $S(p,\ell)$, familiar from the Bethe Ansatz for
spin chains. As we shall see shortly,   $S(p,\ell)$  is determined
by the boundary condition (\ref{BC}) and the contact term
(\ref{I=J}) in the Hamiltonian.  Requiring that the
wave-function is an eigenstate of the contact interaction with the
same eigenvalue leads to the Bethe equation:
\begin{equation}
\left(
e^{i\frac{(p+\ell)}{2}}+e^{-i\frac{(p+\ell)}{2}}-2e^{i\frac{(p-\ell)}{2}}\right)
+ \left(
e^{i\frac{(p+\ell)}{2}}+e^{-i\frac{(p+\ell)}{2}}-2e^{-i\frac{(p-\ell)}{2}}\right)S(p,\ell)
e^{i(p-\ell)}=0 \label{bethe}
\end{equation}
To proceed, we need to impose the boundary condition (\ref{BC}).
It implies
$$
e^{ipI+i\ell (M+1)}+S(p,\ell) e^{i\ell I+ip(M+1)} =e^{ip+i\ell
I}+S(p,\ell) e^{i\ell +ipI}
$$
which can be simplified to
$$
e^{i(p-\ell)(I-1)+i\ell M}+S(p,\ell) e^{ipM} = 1+S(p,\ell)
e^{i(p-\ell)(I-1)}
$$
This equation must hold for any value of $I$.  Simple algebra
gives the two conditions \begin{equation} S=e^{i\ell M}
~~~,~~~e^{i(p+\ell)M}=1\longrightarrow p+\ell=2\pi j/M
\label{p+l}\end{equation} This latter quantization of $p+\ell$ can
also be seen as a result of invariance of (\ref{I<J}) and
(\ref{I=J}) under translating both variables $(I,J)\to (I+M,J+M)$.

The Bethe equation (\ref{bethe}) becomes \begin{equation}\left(
\cos \frac{\pi j}{M} -
e^{i\frac{(p-\ell)}{2}}\right)+(-1)^j\left(\cos\frac{\pi
j}{M}-e^{-i\frac{(p-l)}{2}}\right)e^{i\frac{(p-l)}{2}(2-M)}=0
\end{equation}
This equation is easily solved in the large $M$ limit, which is
the case of most interest to us. The result is
\begin{equation}
\frac{p+\ell}{2}=\frac{\pi j}{M}~~~ \frac{p-\ell}{2}=\frac{\pi n
}{M}+ {\cal O}\left(\frac{1}{M^2}\right)
\end{equation} where $j$ and $n$ are either both even or both odd integers.
 Their sum or difference are therefore always even and are equal to
 two times any integer.  Thus, when $M$ is large,
$$
\left(p,\ell\right)=\frac{2\pi}{M}\left(r,s\right)+{\cal
O}\left(\frac{1}{M}\right)~~~~r,s\in{\cal Z}
$$
and
$$
S=1+{\cal O}\left(\frac{1}{M}\right)
$$
Thus, we see that, in the large $M$ limit, the extension of the
function ${\cal O}_{IJ}$, which was defined only for $I\leq J$, to
all $I$ and $J$, is as a symmetric doubly periodic function. In
this limit, the spectrum of the Hamiltonian is
\begin{equation}
\frac{g_{YM}^2 N}{2M} \left(r^2+s^2\right)
\end{equation}

Note that this matches the string spectrum in the low energy
limit, for the state which contains two oscillators, created from
the sigma model vacuum by $a_r^{\dagger}$ and $a_s^{\dagger}$ and
which has one unit of light-cone momentum, $k=1$.

Now, we consider the operator where two impurities are inserted
into  $k$ chains $A_1\ldots A_M$. It can have the form
\begin{equation}
{\cal O}_{IJ} = {\rm TR}\left( A_1\ldots\Phi_I\ldots \Phi_J\ldots
A_M(A_1\ldots A_M)^{k-1}\right)
\end{equation} or \begin{equation}
{\cal O}_{IJ}={\rm TR}\left(A_1\ldots\Phi_I\ldots A_M(A_1\ldots
A_M)^{k'}A_1\ldots\Phi_J\ldots A_M(A_1\ldots
A_M)^{k-k'-2}\right)\end{equation}

We can simply treat these as a function where $I\leq J$ and both
$I$ and $J$ run from $1$ to $kM$.  Then, the previous discussion
of the case where $k=1$ can be applied here with the only
difference that the integers $(r,s)$ are replaced by the ratios
$\left(\frac{r}{k},\frac{s}{k}\right)$.  The spectrum of the
Hamiltonian is given by
\begin{equation} \frac{g_{YM}^2 N}{2Mk^2}
\left(r^2+s^2\right)
\end{equation}
This matches the noninteracting string spectrum when there are $k$
units of light-cone momentum.  However, all values of $r$ and $s$
are not allowed, but are restricted by a further condition.  To
see this, note that we have not yet taken into account that the
operator ${\cal O}_{IJ}$ is periodic under the shift $(I,J)\to
(I+M,J+M)$, this periodicity follows from cyclicity of the trace.
We see that the wavefunction transforms as
$$
\psi_{I+MJ+M}=\psi_{IJ}e^{2\pi
i\left(\frac{r}{k}+\frac{s}{k}\right)} $$

In order for the wave-functions to have the correct periodicity,
it is necessary that  $\frac{r}{k}+\frac{s}{k}$ is an integer. For
this to be the case, $r$ and $s$ must satisfy the further
condition $r+s=km$ where $m$ is an integer. It is this integer,
$m$, which is identified with the wrapping number in string theory
and the condition that $r+s=km$ then coincides with the
level-matching condition, which poses the same restriction on
string theory states.

The eigenstates of the Hamiltonian are to be used as follows.
First, it is straightforward to show that the contact term in the
Hamiltonian is precisely what is needed to make the Hamiltonian
Hermitian. It does this by cancelling boundary effects in
summations by parts in the inner product, so that $$
<\psi_1|H_0\psi_2>=<H_0\psi_1|\psi_2>
$$
where
$$
<\psi_1|\psi_2> =\sum_{I\leq
J=1}^{kM}\psi^{\dagger}_{1IJ}\psi_{2IJ}
$$
 The operator ${\cal O}_{IJ}$ can be
written as a superposition of eigenstates as
$$
{\cal O}_{IJ}=\sum_E \psi_{IJ}(E){\cal O}(E)
$$
and
$$
{\cal O}(E)=\frac{2}{kM(kM+1)}\sum_{I\leq
J=1}^{kM}\psi_{IJ}^{\dagger}(E){\cal O}_{IJ}
$$

\section{The $k=3$ wavefunctions $w(x,y)$, $v(x,y)$ and $\psi_\pm(x,y)$}

In this appendix we first provide the explicit form of the $k=3$ states
$w(x,y)$ and $v(x,y)$
that are gotten by the repeated action
of the operator $(D-3M-2)$ on the general linear combination
$u(x,y)$, (\ref{u}), of the 9 basis
states (\ref{basis}). The action of $(D-3M-2)$ on $u,\ w$ and $v$ is given in
(\ref{set1}).
\bea
w(x,y)&=&
\theta(y-x)\left[(c_5-c_4)O^{S3}(x,y)-(c_5+4c_6)O^{S3}(y,x)+
(c_4+4c_6)f^{C3}(x,y)\right.\cr
&&\left.-(c_1-c_3-3c_7)f^{C21}(x,y)+(c_1-c_2-3c_7)O^{S21}(x,y)
+(c_3-c_2)D^{12}(x,y)\right.\cr
&&\left.+(c_4-c_5)T^{111}(x,y)\right]\cr
&+&\theta(x-y)\left[-(c_5+4c_6)O^{S3}(x,y)+(c_5-c_4)O^{S3}(y,x)\right.\cr
&&\left.+(c_4+4c_6)f^{C3}(x,y)
-(c_1-c_3-3c_7)f^{C21}(x,y)+(c_1-c_2-3c_7)O^{S21}(x,y)\right.\cr
&&\left.+(c_3-c_2)D^{12}(x,y)+(c_4-c_5)T^{111}(x,y)\right]
\label{w}
\eea
\bea
v(x,y)&=&\theta(y-x)\left[(2c_1-c_3-c_2-6c_7)O^{S3}(x,y)+(5c_2+3c_7-c_1-4c_3)O^{S3}(y,x)\right.\cr
&&\left.+
(5c_3-4c_2-c_1+3c_7)f^{C3}(x,y)
+(5c_4-4c_5+4c_6)f^{C21}(x,y)\right.\cr
&&\left.+(5c_5-4c_4+4c_6)O^{S21}(x,y)+(c_4+c_5+8c_6)D^{12}(x,y)\right.\cr
&&\left.
+(c_2+c_3+6c_7-2c_1)T^{111}(x,y)\right]\cr
&+&\theta(x-y)\left[(5c_2+3c_7-c_1-4c_3)O^{S3}(x,y)+(2c_1-c_2-c_3-6c_7)O^{S3}(y,x)\right.\cr
&&\left.+
(5c_3-4c_2-c_1+3c_7)f^{C3}(x,y)
+(5c_4-4c_5+4c_6)f^{C21}(x,y)\right.\cr
&&\left.+(5c_5-4c_4+4c_6)O^{S21}(x,y)+(c_4+c_5+8c_6)D^{12}(x,y)\right.\cr
&&\left.
+(c_2+c_3+6c_7-2c_1)T^{111}(x,y)\right]
\label{v}
\eea

The states $\psi_\pm(x,y)$ that diagonalize the operator $(D-3M-2)$
and that, consequently, have a well defined anomalous dimension, in
terms of the original coefficients of $u(x,y)$, (\ref{u}), are given by
\bea
\psi_+(x,y)
&=&\theta(y-x)\left[(2c_1-c_2-c_3-3c_4+3 c_5-6c_7)O^{S3}(x,y)\right.\cr
&&\left.
+(-c_1 +5c_2-4c_3-3 c_5 -12 c_6+3c_7)O^{S3}(y,x)\right.\cr
&&\left.+
(-c_1-4c_2+5c_3+3c_4+12c_6+3c_7)f^{C3}(x,y)\right.\cr
&&\left.
+(-3c_1+3c_3+5c_4-4c_5+4c_6+9c_7)f^{C21}(x,y)\right.\cr
&&\left.
+(3c_1-3c_2-4c_4+ 5c_5+4c_6-9c_7)O^{S21}(x,y)\right.\cr
&&\left.+(-3c_2+3c_3+c_4+c_5+8c_6)D^{12}(x,y)\right.\cr
&&\left.
+(-2c_1+c_2+c_3+3c_4-3c_5+6c_7)T^{111}(x,y)\right]\cr
&+&\theta(x-y)
\left[(-c_1 +5c_2-4c_3-3 c_5 -12 c_6+3c_7)O^{S3}(x,y)\right.\cr
&&\left.
+(2c_1-c_2-c_3-3c_4+3 c_5-6c_7)O^{S3}(y,x)\right.\cr
&&\left.+
(-c_1-4c_2+5c_3+3c_4+12c_6+3c_7)f^{C3}(x,y)\right.\cr
&&\left.
+(-3c_1+3c_3+5c_4-4c_5+4c_6+9c_7)f^{C21}(x,y)\right.\cr
&&\left.
+(3c_1-3c_2-4c_4+ 5c_5+4c_6-9c_7)O^{S21}(x,y)\right.\cr
&&\left.+(-3c_2+3c_3+c_4+c_5+8c_6)D^{12}(x,y)\right.\cr
&&\left.
+(-2c_1+c_2+c_3+3c_4-3c_5+6c_7)T^{111}(x,y)\right]
\label{psip}
\eea
\bea
\psi_-(x,y)
&=&\theta(y-x)\left[(-2c_1+c_2+c_3-3c_4+3 c_5+6c_7)O^{S3}(x,y)\right.\cr
&&\left.
+(c_1 -5c_2+4c_3-3 c_5 -12 c_6-3c_7)O^{S3}(y,x)\right.\cr
&&\left.+
(c_1+4c_2-5c_3+3c_4+12c_6-3c_7)f^{C3}(x,y)\right.\cr
&&\left.
+(-3c_1+3c_3-5c_4+4c_5-4c_6+9c_7)f^{C21}(x,y)\right.\cr
&&\left.
+(3c_1-3c_2+4c_4- 5c_5-4c_6-9c_7)O^{S21}(x,y)\right.\cr
&&\left.+(-3c_2+3c_3-c_4-c_5-8c_6)D^{12}(x,y)\right.\cr
&&\left.
+(2c_1-c_2-c_3+3c_4-3c_5-6c_7)T^{111}(x,y)\right]\cr
&+&\theta(x-y)\left[(c_1 -5c_2+4c_3-3 c_5 -12 c_6-3c_7)O^{S3}(x,y)\right.\cr
&&\left.
+(-2c_1+c_2+c_3-3c_4+3 c_5+6c_7)O^{S3}(y,x)\right.\cr
&&\left.+
(c_1+4c_2-5c_3+3c_4+12c_6-3c_7)f^{C3}(x,y)\right.\cr
&&\left.
+(-3c_1+3c_3-5c_4+4c_5-4c_6+9c_7)f^{C21}(x,y)\right.\cr
&&\left.
+(3c_1-3c_2+4c_4- 5c_5-4c_6-9c_7)O^{S21}(x,y)\right.\cr
&&\left.+(-3c_2+3c_3-c_4-c_5-8c_6)D^{12}(x,y)\right.\cr
&&\left.
+(2c_1-c_2-c_3+3c_4-3c_5-6c_7)T^{111}(x,y)\right]
\label{psim}
\eea
Denoting by $\psi^i_\pm$ $i=1,\dots,7$ the coefficients of the 7
operators in $\psi_\pm(x,y)$,
it is easy to find 5 independent relations between these coefficients
both in
$\psi_+(x,y)$ and in $\psi_-(x,y)$. For $y\ge x$ one has
\bea
&&\psi^1_\pm+\psi^2_\pm+\psi^3_\pm=0\cr
&&\psi^4_\pm+\psi^5_\pm-\psi^6_\pm=0\cr
&&\psi^1_\pm+\psi^7_\pm=0\cr
&&4\psi^1_\pm-\psi^3_\pm\pm3\psi^4_\pm=0\cr
&&4\psi^1_\pm-\psi^2_\pm\mp3\psi^5_\pm=0
\eea
for $y\le x$ it is sufficient to exchange $1\leftrightarrow 2$ in the above.
Therefore the number of independent states of this form is 4. These are
the states with
period 6 that should have corrections to all orders in the string loop
expansion, whereas
the other 5 states needed to complete the Hamiltonian eigenfunctions in
the $k=3$ case,
$u^i(x,y)$ $i=1,\dots,5$, are given in the text and do not have corrections
beyond the planar level.


\begin{thebibliography}{99}

%\cite{Maldacena:1997re}
\bibitem{Maldacena:1997re}
J.~M.~Maldacena,
``The large N limit of superconformal field theories and supergravity,''
Adv.\ Theor.\ Math.\ Phys.\  {\bf 2}, 231 (1998)
[Int.\ J.\ Theor.\ Phys.\  {\bf 38}, 1113 (1999)]
[arXiv:hep-th/9711200].
%%CITATION = HEP-TH 9711200;%%

%\cite{Gubser:1998bc}
\bibitem{Gubser:1998bc}
S.~S.~Gubser, I.~R.~Klebanov and A.~M.~Polyakov,
``Gauge theory correlators from non-critical string theory,''
Phys.\ Lett.\ B {\bf 428}, 105 (1998)
[arXiv:hep-th/9802109].
%%CITATION = HEP-TH 9802109;%%

%\cite{Witten:1998qj}
\bibitem{Witten:1998qj}
E.~Witten,
``Anti-de Sitter space and holography,''
Adv.\ Theor.\ Math.\ Phys.\  {\bf 2}, 253 (1998)
[arXiv:hep-th/9802150].
%%CITATION = HEP-TH 9802150;%%

%\cite{Lee:1998bx}
\bibitem{Lee:1998bx}
S.~M.~Lee, S.~Minwalla, M.~Rangamani and N.~Seiberg, ``Three-point
functions of chiral operators in D = 4, N = 4 SYM at  large N,''
Adv.\ Theor.\ Math.\ Phys.\  {\bf 2}, 697 (1998)
[arXiv:hep-th/9806074].
%%CITATION = HEP-TH 9806074;%%

%\cite{Freedman:1998tz}
\bibitem{Freedman:1998tz}
D.~Z.~Freedman, S.~D.~Mathur, A.~Matusis and L.~Rastelli,
``Correlation functions in the CFT($d$)/AdS($d+1$)
correspondence,'' Nucl.\ Phys.\ B {\bf 546}, 96 (1999)
[arXiv:hep-th/9804058].
%%CITATION = HEP-TH 9804058;%%

%\cite{Chalmers:1998xr}
\bibitem{Chalmers:1998xr}
G.~Chalmers, H.~Nastase, K.~Schalm and R.~Siebelink, ``R-current
correlators in N = 4 super Yang-Mills theory from anti-de  Sitter
%supergravity,''
Nucl.\ Phys.\ B {\bf 540}, 247 (1999) [arXiv:hep-th/9805105].
%%CITATION = HEP-TH 9805105;%%

%\cite{Erickson:2000af}
\bibitem{Erickson:2000af}
J.~K.~Erickson, G.~W.~Semenoff and K.~Zarembo, ``Wilson loops in N
= 4 supersymmetric Yang-Mills theory,'' Nucl.\ Phys.\ B {\bf 582},
155 (2000) [arXiv:hep-th/0003055].
%%CITATION = HEP-TH 0003055;%%



%\cite{Drukker:2000rr}
\bibitem{Drukker:2000rr}
N.~Drukker and D.~J.~Gross, ``An exact prediction of N = 4
supersymmetryM theory for string theory,'' J.\ Math.\ Phys.\  {\bf
42}, 2896 (2001) [arXiv:hep-th/0010274].
%%CITATION = HEP-TH 0010274;%%

%\cite{Semenoff:2001xp}
\bibitem{Semenoff:2001xp}
G.~W.~Semenoff and K.~Zarembo, ``More exact predictions of SUSYM
for string theory,'' Nucl.\ Phys.\ B {\bf 616}, 34 (2001)
[arXiv:hep-th/0106015].
%%CITATION = HEP-TH 0106015;%%

%\cite{Semenoff:2002kk}
\bibitem{Semenoff:2002kk}
G.~W.~Semenoff and K.~Zarembo, ``Wilson loops in supersymmetric
Yang-Mills theory theory: From weak to strong coupling,'' Nucl.\
Phys.\ Proc.\ Suppl.\  {\bf 108}, 106 (2002)
[arXiv:hep-th/0202156].
%%CITATION = HEP-TH 0202156;%%

%\cite{Blau:2002dy}
\bibitem{Blau:2002dy}
M.~Blau, J.~Figueroa-O'Farrill, C.~Hull and G.~Papadopoulos,
``Penrose limits and maximal supersymmetry,''
Class.\ Quant.\ Grav.\  {\bf 19}, L87 (2002)
[arXiv:hep-th/0201081].
%%CITATION = HEP-TH 0201081;%%
%\cite{Blau:2001ne}
%\bibitem{Blau:2001ne}
M.~Blau, J.~Figueroa-O'Farrill, C.~Hull and G.~Papadopoulos,
``A new maximally supersymmetric background of IIB superstring theory,''
JHEP {\bf 0201}, 047 (2002)
[arXiv:hep-th/0110242].
%%CITATION = HEP-TH 0110242;%%

%\cite{Gueven:2000ru}
\bibitem{Gueven:2000ru}
R.~Gueven,
``Plane wave limits and T-duality,''
Phys.\ Lett.\ B {\bf 482}, 255 (2000)
[arXiv:hep-th/0005061].
%%CITATION = HEP-TH 0005061;%%
%\cite{Metsaev:2001bj}
\bibitem{Metsaev:2001bj}
R.~Metsaev,
``Type IIB Green-Schwarz superstring in plane wave Ramond-Ramond  background,''
Nucl.\ Phys.\ B {\bf 625}, 70 (2002)
[arXiv:hep-th/0112044].
%%CITATION = HEP-TH 0112044;%%


\bibitem{Berenstein:2002jq}
D.~Berenstein, J.~M.~Maldacena and H.~Nastase,
``Strings in flat space and pp waves from N = 4 super Yang Mills,''
JHEP {\bf 0204}, 013 (2002)
[arXiv:hep-th/0202021].
%%CITATION = HEP-TH 0202021;%%

%\cite{'tHooft:1973jz}
\bibitem{'tHooft:1973jz}
G.~'t Hooft,
``A Planar Diagram Theory For Strong Interactions,''
Nucl.\ Phys.\ B {\bf 72}, 461 (1974).
%%CITATION = NUPHA,B72,461;%%

%\cite{Gross:2002su}
\bibitem{Gross:2002su}
D.~J.~Gross, A.~Mikhailov and R.~Roiban,
``Operators with large R charge in N = 4 Yang-Mills theory,''
Annals Phys.\  {\bf 301}, 31 (2002)
[arXiv:hep-th/0205066].



\bibitem{Kristjansen:2002bb}
C.~Kristjansen, J.~Plefka, G.~W.~Semenoff and M.~Staudacher,
``A new double-scaling limit of N = 4 super Yang-Mills theory and
PP-wave  strings,''
Nucl.\ Phys.\ B {\bf 643}, 3 (2002)
[arXiv:hep-th/0205033].
%%CITATION = HEP-TH 0205033;%%

\bibitem{Constable:2002hw}
N.~R.~Constable, D.~Z.~Freedman, M.~Headrick,
S.~Minwalla, L.~Motl, A.~Postnikov and W.~Skiba,
``PP-wave string interactions from perturbative Yang-Mills theory,''
JHEP {\bf 0207}, 017 (2002)
[arXiv:hep-th/0205089].
%%CITATION = HEP-TH 0205089;%%

%\cite{Bianchi:2002rw}
\bibitem{Bianchi:2002rw}
M.~Bianchi, B.~Eden, G.~Rossi and Y.~S.~Stanev,
``On operator mixing in N = 4 SYM,''
Nucl.\ Phys.\ B {\bf 646}, 69 (2002)
[arXiv:hep-th/0205321].
%%CITATION = HEP-TH 0205321;%%

%\cite{Beisert:2002bb}
\bibitem{Beisert:2002bb}
N.~Beisert, C.~Kristjansen, J.~Plefka, G.~W.~Semenoff and M.~Staudacher,
``BMN correlators and operator mixing in N = 4 super Yang-Mills theory,''
Nucl.\ Phys.\ B {\bf 650}, 125 (2003)
[arXiv:hep-th/0208178].
%%CITATION = HEP-TH 0208178;%%

%\cite{Constable:2002vq}
\bibitem{Constable:2002vq}
N.~R.~Constable, D.~Z.~Freedman, M.~Headrick and S.~Minwalla,
``Operator mixing and the BMN correspondence,''
JHEP {\bf 0210}, 068 (2002)
[arXiv:hep-th/0209002].
%%CITATION = HEP-TH 0209002;%%

%\cite{Spradlin:2002ar}
\bibitem{Spradlin:2002ar}
M.~Spradlin and A.~Volovich,
``Superstring interactions in a pp-wave background,''
Phys.\ Rev.\ D {\bf 66}, 086004 (2002)
[arXiv:hep-th/0204146].
%%CITATION = HEP-TH 0204146;%%


%\cite{Chu:2002pd}
\bibitem{Chu:2002pd}
C.~S.~Chu, V.~V.~Khoze and G.~Travaglini,
``Three-point functions in N = 4 Yang-Mills theory and pp-waves,''
JHEP {\bf 0206}, 011 (2002)
[arXiv:hep-th/0206005].
%%CITATION = HEP-TH 0206005;%%


%\cite{Spradlin:2002rv}
\bibitem{Spradlin:2002rv}
M.~Spradlin and A.~Volovich,
``Superstring interactions in a pp-wave background. II,''
JHEP {\bf 0301}, 036 (2003)
[arXiv:hep-th/0206073].
%%CITATION = HEP-TH 0206073;%%

%\cite{Chu:2002qj}
\bibitem{Chu:2002qj}
C.~S.~Chu, V.~V.~Khoze and G.~Travaglini,
``pp-wave string interactions from n-point correlators of BMN operators,''
JHEP {\bf 0209}, 054 (2002)
[arXiv:hep-th/0206167].
%%CITATION = HEP-TH 0206167;%%

%\cite{Chu:2002eu}
\bibitem{Chu:2002eu}
C.~S.~Chu, V.~V.~Khoze, M.~Petrini, R.~Russo and A.~Tanzini,
``A note on string interaction on the pp-wave background,''
Class.\ Quant.\ Grav.\  {\bf 21}, 1999 (2004)
[arXiv:hep-th/0208148].
%%CITATION = HEP-TH 0208148;%%

%\cite{Pankiewicz:2002gs}
\bibitem{Pankiewicz:2002gs}
A.~Pankiewicz,
``More comments on superstring interactions in the pp-wave background,''
JHEP {\bf 0209}, 056 (2002)
[arXiv:hep-th/0208209].
%%CITATION = HEP-TH 0208209;%%

%\cite{Pankiewicz:2002tg}
\bibitem{Pankiewicz:2002tg}
A.~Pankiewicz and B.~.~J.~Stefanski,
``pp-wave light-cone superstring field theory,''
Nucl.\ Phys.\ B {\bf 657}, 79 (2003)
[arXiv:hep-th/0210246].
%%CITATION = HEP-TH 0210246;%%


%\cite{Chu:2002wj}
\bibitem{Chu:2002wj}
C.~S.~Chu, M.~Petrini, R.~Russo and A.~Tanzini,
``String interactions and discrete symmetries of the pp-wave background,''
Class.\ Quant.\ Grav.\  {\bf 20}, S457 (2003)
[arXiv:hep-th/0211188].
%%CITATION = HEP-TH 0211188;%%

%%\cite{He:2002zu}
\bibitem{He:2002zu}
Y.~H.~He, J.~H.~Schwarz, M.~Spradlin and A.~Volovich,
``Explicit formulas for Neumann coefficients in the plane-wave geometry,''
Phys.\ Rev.\ D {\bf 67}, 086005 (2003)
[arXiv:hep-th/0211198].
%%CITATION = HEP-TH 0211198;%%

\bibitem{Gutjahr:2004dv}
P.~Gutjahr and A.~Pankiewicz,
``New aspects of the BMN correspondence beyond the planar limit,''
arXiv:hep-th/0407098.
%%CITATION = HEP-TH 0407098;%%

\bibitem{PandoZayas:2002hh}
L.~A.~Pando Zayas and D.~Vaman,
``Strings in RR plane wave background at finite temperature,''
Phys.\ Rev.\ D {\bf 67}, 106006 (2003)
[arXiv:hep-th/0208066],
B.~R.~Greene, K.~Schalm and G.~Shiu,
``On the Hagedorn behaviour of pp-wave strings and N = 4 SYM theory at finite
R-charge density,''
Nucl.\ Phys.\ B {\bf 652} (2003) 105
[arXiv:hep-th/0208163],
Y.~Sugawara,
``Thermal amplitudes in DLCQ superstrings on pp-waves,''
Nucl.\ Phys.\ B {\bf 650}, 75 (2003)
[arXiv:hep-th/0209145],
R.~C.~Brower, D.~A.~Lowe and C.~I.~Tan,
``Hagedorn transition for strings on pp-waves and tori with chemical
potentials,''
Nucl.\ Phys.\ B {\bf 652}, 127 (2003)
[arXiv:hep-th/0211201],
G.~Grignani, M.~Orselli, G.~W.~Semenoff and D.~Trancanelli,
``The superstring Hagedorn temperature in a pp-wave background,''
JHEP {\bf 0306}, 006 (2003)
[arXiv:hep-th/0301186],
O.~Aharony, J.~Marsano, S.~Minwalla, K.~Papadodimas and M.~Van Raamsdonk,
``The Hagedorn / deconfinement phase transition in weakly coupled large N gauge
theories,''
arXiv:hep-th/0310285.
%%CITATION = HEP-TH 0310285;%%

\bibitem{Mukhi:2002ck}
S. Mukhi, M. Rangamani and E. Verlinde,
``Strings from Quivers, Membranes from Moose''
JHEP {\bf 0205}, 023 2002
[arXiv:hep-th/020204147]


\bibitem{Bertolini:2002nr}
M.~Bertolini, J.~de Boer, T.~Harmark, E.~Imeroni and N.~A.~Obers,
``Gauge theory description of compactified pp-waves,''
JHEP {\bf 0301}, 016 (2003)
[arXiv:hep-th/0209201].
%%CITATION = HEP-TH 0209201;%%

\bibitem{Douglas:1996sw}
M.~R.~Douglas and G.~W.~Moore,
``D-branes, Quivers, and ALE Instantons,''
arXiv:hep-th/9603167.
%%CITATION = HEP-TH 9603167;%%

\bibitem{Bershadsky:1998cb}
M.~Bershadsky and A.~Johansen, 
``Large {$N$} limit of orbifold field theories,'' 
{\em Nucl. Phys.} {\bf B536} (1998) 141--148,
\href{http://arXiv.org/abs/hep-th/9803249}{{\tt hep-th/9803249}}.
%%CITATION = HEP-TH 9803249;%%.

%\cite{Alishahiha:2002ev}
\bibitem{Alishahiha:2002ev}
M.~Alishahiha and M.~M.~Sheikh-Jabbari,
``The pp-wave limits of orbifolded AdS(5) x S(5),''
Phys.\ Lett.\ B {\bf 535}, 328 (2002)
[arXiv:hep-th/0203018].
%%CITATION = HEP-TH 0203018;%%



%\cite{Minahan:2002ve}
\bibitem{Minahan:2002ve}
J.~A.~Minahan and K.~Zarembo,
``The Bethe-ansatz for N = 4 super Yang-Mills,''
JHEP {\bf 0303}, 013 (2003) [arXiv:hep-th/0212208].
%%CITATION = HEP-TH 0212208;%%

\bibitem{Beisert:2002ff}
N.~Beisert, C.~Kristjansen, J.~Plefka and M.~Staudacher,
``BMN gauge theory as a quantum mechanical system,''
Phys.\ Lett.\ B {\bf 558}, 229 (2003)
[arXiv:hep-th/0212269].
%%CITATION = HEP-TH 0212269;%%

\bibitem{Beisert:2003tq}
N.~Beisert, C.~Kristjansen and M.~Staudacher,
``The dilatation operator of N = 4 super Yang-Mills theory,''
Nucl.\ Phys.\ B {\bf 664}, 131 (2003)
[arXiv:hep-th/0303060].
%%CITATION = HEP-TH 0303060;%%

\bibitem{Dobashi:2004nm}
Dobashi and T.~Yoneya,
``Resolving the holography in the plane-wave limit of AdS/CFT correspondence,''
arXiv:hep-th/0406225.
%%CITATION = HEP-TH 0406225;%%

%\cite{Lee:2004cq}
\bibitem{Lee:2004cq}
S.~Lee and R.~Russo,
``Holographic cubic vertex in the pp-wave,''
arXiv:hep-th/0409261.
%%CITATION = HEP-TH 0409261;%%







\end{thebibliography}
\end{document}